\documentclass[10pt,twocolumn]{article}

\usepackage{newtxtext,newtxmath}

\usepackage{graphicx}

\usepackage[letterpaper,margin=1in]{geometry}

\renewenvironment{abstract}
	{\quotation}
	{\endquotation}

\date{}

\makeatletter
\renewcommand{\fnum@figure}{\textbf{Figure \thefigure}}
\renewcommand{\fnum@table}{\textbf{Table \thetable}}
\makeatother

\usepackage{scicite}

\usepackage{url}

\usepackage{matlab-prettifier}
\newcommand{\matlab}[1]{\lstinline[style=Matlab-editor]!#1!}
\usepackage{newtxtext,newtxmath}
\usepackage[superscript,biblabel]{cite}
\usepackage{chemformula} 
\usepackage[T1]{fontenc} 
\usepackage{caption}
\usepackage[mathlines]{lineno} 
\usepackage{lipsum}
\usepackage{placeins}
\usepackage[acronyms,shortcuts]{glossaries}
\usepackage{textcomp}
\usepackage{gensymb}
\usepackage{float}
\usepackage{braket}
\usepackage{cleveref}
\usepackage{nameref}

\def\scititle{
	\textit{In situ} Learning-Based Spin Engineering of Pulsed Dynamic Nuclear Polarization
}
\title{\bfseries \boldmath \scititle}

\author{
	Filip V. Jensen$^{1}$, 
	José P. Carvalho$^{1}$,
	Nino Wili$^{1}$,
	Asbj{\o}rn Holk Thomsen$^{2}$,
	David L. Goodwin$^{1}$,\\
	Lukas Trottner$^{3}$,
	Claudia Strauch$^{4}$,
	Anders Bodholt Nielsen$^{1\ast}$,
	Niels Chr. Nielsen$^{1\ast}$\and
	\small$^{1}$Interdisciplinary Nanoscience Center (iNANO) and Department of Chemistry, Aarhus University, \and \small Gustav Wieds Vej 14, DK-8000 Aarhus C, Denmark.\and 
	\small$^{2}$Department of Mathematics, Aarhus University, Ny Munkegade 118, DK-8000 Aarhus C, Denmark.\and 
	\small$^{3}$Institute for Stochastics and Applications, University of Stuttgart, Wankelstrasse 5, 70563 Stuttgart, Germany.\and 
	\small$^{4}$Institut f\"ur Mathematik, Ruprecht-Karls-Universit\"at Heidelberg, Mathematikon, \\\small Im Neuenheimer Feld 205, 69120 Heidelberg, Germany.\\
	\small$^\ast$Corresponding authors. Email: ncn@chem.au.dk abn@chem.au.dk
}

\begin{document} 
\twocolumn[
\begin{@twocolumnfalse}

\maketitle


\begin{abstract} 
Pulsed Dynamic Nuclear Polarization (DNP) is currently receiving substantial interest as a means to enhance the sensitivity of nuclear magnetic resonance (NMR) and magnetic resonance imaging (MRI) by orders of magnitude. It has also received much attention as a central ingredient in many modalities of electron spin-involved quantum sensing. Relative to spin engineering associated with NMR, the design of efficient pulsed DNP experiments with a broad experimental scope are challenged by large electron-nuclear spin systems, large electron spin-involved interactions, and instrumental non-idealities and limitations. All of this may challenge traditional NMR-like theoretical and numerical pulse sequence engineering. Exploiting state-of-the-art instrumentation and taking advantage of the high sensitivity of DNP relative to NMR, we here demonstrate the use of combinations of Bayesian machine learning methods and constrained random walk procedures to design pulse sequences \textit{in situ}, by experiments, directly on the spin systems responding to spectrometer instructions. For trityl and nitroxide samples, it is demonstrated that efficient broadband DNP pulse sequences can be designed \textit{in situ} with experimental protocols benchmarked against \textit{in silico} analogs.
\end{abstract}

\end{@twocolumnfalse}
]
\section{Introduction}
\label{sec:Introduction}

Coherent spectroscopy has undergone a tremendous evolution in an increasing span of disciplines, since the pioneering magnetic resonance experiments in the 1940's.\cite{Bloch,Hahn,Ramsey_1950,Hartmann_photon_echo_1964,Freed_coherent_EPR,abragam1961principles,Ernst-book,Jeschke_book} This has been driven by an intricate interplay between the development of increasingly advanced instrumentation and the exploitation of powerful analytical and numerical methods to design pulsed experiments while taking advantage of the gradually improving instrumentation. This process has been further fueled by emerging needs for information on the structures and dynamics of molecular systems in many disciplines. A prime example is nuclear magnetic resonance (NMR) which may be considered the “easy” case due to weak spin interactions and the associated long coherence lifetimes available for manipulation of nuclear spin systems. In the context of magnetic resonance, advanced pulse engineering is, to an increasing extent, being applied to methods involving electron spins, such as paramagnetic NMR, electron paramagnetic resonance (EPR), and dynamic nuclear polarization (DNP). DNP bridges NMR and EPR by transferring polarization from highly polarized electron spins to nuclear spins to boost the
sensitivity by orders of magnitude and diversifying spectral information. The challenges encountered in EPR and DNP are the large electron spin interactions, fast relaxation, and the mere size of the electron-nuclear spin systems, which may limit current experiment design principles and instrumental realization of optimal experiments.

Several design strategies may be pursued, including analytical (e.g., effective Hamiltonian\cite{AHT,scBCH,EEHT,EEHT2,shankar,SSV-EHT,SVEHT_EEHT}), numerical (e.g., optimal control\cite{grape,MaximovOC,SIMPSONOC,MaximovSmoothing,OCKuprov,OC_DNP}), and experimental (e.g., feedback control\cite{GOODWIN_feedback_epr,Marshall_DNP_insitu}) methods. All of these have advantages and disadvantages that depend heavily on the complexity of the spin systems to be addressed, including the number of spins, interaction strengths, coherence lifetimes, and instrumental capabilities. Realization of pulsed DNP, e.g., in the context of quantum sensing, definitely provides an example of a system much more complicated than typical NMR systems. An unpaired electron spin may have huge interaction strength compared to nuclear spins, implying that typical electron-nuclear spin systems may consist of hundreds to thousands of spins, the coherence lifetimes are typically very short, and the instrumentation may be challenged by insufficient microwave (MW) irradiation power, large MW field inhomogeneity and pulse transients, and insufficient time resolution. In such cases, it is relevant to further explore and potentially reconsider the experiment design process. Both analytical effective-Hamiltonian-based and numerical non-linear optimization or optimal control methods are typically limited in the size and complexity of the spin systems that can be handled. In the latter case, simply due to insufficient capacity to store and operate with density matrices with dimensions scaling exponentially with the number of spins. The most obvious drawback of operating with too small spin systems is that they may be inaccurate or not fully represent the spin systems manipulated and observed by the spectrometer. In this view, it may be interesting to reverse the situation, \textit{i.e.} letting the spectrometer form the basis for an experiment-driven pulse sequence design procedure by  reporting on the spin system for a representative molecular object, as recently explored in combination with optimal control in the context of EPR\cite{GOODWIN_feedback_epr} and DNP.\cite{Marshall_DNP_insitu} Accordingly, a feedback-loop setup may be used to develop experimental methods that respond optimally to the actual type of spin system and  experimental conditions. Realizing \textit{in situ} experiment design obviously calls for high experimental sensitivity and fast repeatability. This may be feasible for pulsed DNP. The challenge remaining is that experimental observables are typically much less detailed than the information content used in numerical gradient-based\cite{grape,SIMPSONOC} or Krotov-type\cite{MaximovOC,MaximovSmoothing} optimal control procedures.

Taking DNP as a guinea pig to explore large-spin-system \textit{in situ} pulse sequence engineering, it is relevant to have ($i$) availability to front-end instrumentation, ($ii$) state-of-the-art pulse sequences, and ($iii$) molecular systems suitable for testing and representative for targeted molecular systems.
Based on pulsed EPR and DNP instrumentation following and extending the protocol of Doll et al,\cite{Doll:2017wb} we have recently developed broadband DNP pulse sequences such as BEAM,\cite{BEAM} PLATO,\cite{PLATO_adv} and cRW-OPT\cite{rCW_EEHT_DNP} which through the combined use of exact effective Hamiltonian theory (EEHT),\cite{EEHT,EEHT2} single-spin-vector effective Hamiltonian theory (SSV-EHT),\cite{shankar,SSV-EHT,SVEHT_EEHT}  constrained random walk,\cite{rCW_EEHT_DNP} and non-linear optimization and optimal control\cite{OC_DNP,carvalho2026} have enabled realization of increasingly strong performance DNP in terms of the range of electron-spin offsets (width of the EPR line) covered. We note that chirped-pulse ENDOR\cite{Chirp_ENDOR_mr-6-33-2025} and, in the context of coherent spectroscopy, DNP echoes\cite{DNP_echo} have also been developed, demonstrating the capability of the setup. The validity of such investigations as a test bench for experimental-driven optimization is reinforced by a remarkably good correspondence between theoretical-numerical  and experimental data for well-characterized trityl-glycerol-water samples.

In this framework, we here investigate the use of a modern machine-learning–based Bayesian probabilistic optimization approach to develop pulse sequences directly on the spectrometer to address the large electron-nuclear spin system and experimental conditions in the design process of DNP experiments. This method is chosen to exploit the signal intensity as the only directly available feedback information differing from the most typical full-density-operator information used in numerical gradient- or Lagrangian-based optimal control optimization methods.\cite{grape,MaximovOC,SIMPSONOC,OCKuprov,OC_DNP}
By having a huge nuclear spin system involved as the receiver of the transfer process with direct incorporation of learning elements, our approach also sets itself apart from
earlier EPR \cite{GOODWIN_feedback_epr}  and DNP chopped random basis (CRAB)  optimal control  \cite{Marshall_DNP_insitu} feedback-loop approaches. 

The Bayesian method works on black-box functions, meaning that it can be applied as long as the objective (e.g., a measurable polarization transfer efficiency) can be evaluated. In addition to this, it is designed for objective functions (which in our case depend on amplitudes and phases of control pulses) that are expensive to evaluate. These attributes make it highly enticing for implementation in the context of pulsed experiments. Since Bayesian optimization does not have any gradients to guide the optimization, it can be beneficial to instead constrain the search space of the control variables. This allows a larger number of variables, in this case MW pulses, to be efficiently optimized. In our case this constraint is a resonance condition derived from a constrained random walk effective Hamiltonian model \cite{rCW_EEHT_DNP}. The performance of the Bayesian optimization was tested for both numerical and experimental evaluations in the context of electron-offset compensating pulsed DNP for trityl and TEMPO nitroxide samples.

\section{Results}
\subsection{Bayesian \textit{in situ} pulse sequence design}
\label{sec:Results}

\begin{figure*}[ht]
    \centering
    \includegraphics[width=\textwidth]{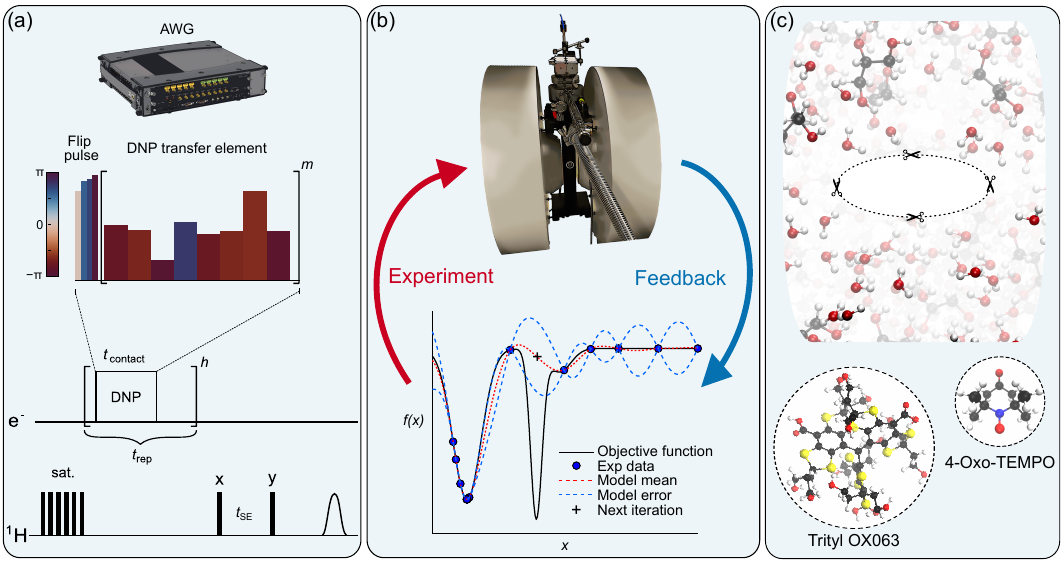}
    \caption{Schematic representation of the experimental setup used for \textit{in situ} Bayesian development of DNP pulse sequences. a) The generation of amplitude and phase modulated pulses for an initial flip (or excitation) pulse and a periodic DNP transfer element in combination mediating  electron to proton polarization transfer. The pulse sequences are generated using an AWG. The overall pulse program contains the initial flip pulse and the DNP transfer element at the MW channel for polarization transfer and an RF channel containing initial saturation pulses and a solid echo for proton signal detection.  b) Part of the spectrometer magnet and the ENDOR/EPR probe for MW/RF pulsing and detection of signals from an inserted frozen radical-containing sample. The spectrometer is controlled by a Bayesian optimization algorithm in a feedback-loop setup receiving response from the experiment and outputs the control variables for the next experiment. c) The sample is a solution containing in our example  trityl OX063 or 4-Oxo-TEMPO radicals.}
    \label{fig:Setup}
\end{figure*}

\textit{In situ} experiment design involves a close interplay between the controlling optimization software, the instrument and its operating system, and the sample providing the response signal to evaluate. This forms the vehicle for iterative pulse sequence development in an artificial intelligence feedback-loop setup. The core principle of our proposed setup is illustrated in Figure \ref{fig:Setup},
involving instrumentation, pulse sequence, learning based feedback control, and a sample with a representative spin system. First, it is crucial that the DNP instrument, in a stand-alone setup, is capable of realizing advanced DNP pulse sequences to drive polarization efficiently from an unpaired electron spin to surrounding protons. A key ingredient in the process is a fast AWG to transmit shaped pulse sequences to the sample as illustrated in Fig. \ref{fig:Setup}a, with the overall pulse sequence sketched in the lower part of the panel. Here, it should be noted that what we optimize is the DNP element (DNP box in Fig. \ref{fig:Setup}a) with the experiment additionally involving repeated pumping of polarization to, and via spin diffusion through, the large nuclear spin system. The performance of the system is clearly important for our exploration of \textit{in situ} optimization vs. state-of-the-art pulsed DNP experiments, designed analytically and/or numerically. One may argue that, in later more general practical applications, the \textit{in situ} design protocol may be even more useful in exploiting instrumental setups, e.g., with lower power and time digitization. The DNP experiment is invoked on a sample in the spectrometer (i.e., in EPR/ENDOR resonator located in the magnet), providing feedback to the Bayesian optimization, as illustrated in Fig. \ref{fig:Setup}b. The samples, and thereby their "real-life" representative spin systems, used in this study are illustrated in Fig. \ref{fig:Setup}c . This involves  Trityl OX063 or 4-Oxo-TEMPO radicals in a deuterated-glycerol-H$_2$O-D$_2$O solutions frozen at 80 K. 

The DNP spectrometer is controlled by Matlab,\cite{MATLAB:2024b} which readily enables a combination with Bayesian optimization protocols also available in Matlab. The Bayesian optimization operates directly on the experimental feedback DNP signal. For the narrow-EPR-line trityl sample, it utilizes multiple measurements taken at different electron spin offsets, whereas for the broad-EPR-line  nitroxide sample, it utilizes a single measurement taken at the carrier frequency. The method operates through learning, without guidance from gradients or other constraints. This may require a relatively large number of evaluations to reach optimal performance depending on the number of variables in the optimization. In practical applications, as demonstrated below, it appears advisable to use a relatively low number of variables. In the interest of fast optimization, we, in addition to this Bayesian stand-alone approach, explored combinations with constrained random walk (cRW)\cite{rCW_EEHT_DNP}.  This method provides a fast way to generate sequences fulfilling a DNP resonance condition ({\it vide infra}) to limit the search space of the  Bayesian optimization. The latter approach may be useful in cases where dominant small-spin-system interactions are expected to be present, releasing the focus of the Bayesian optimization from the large-spin-system network of interactions. Details on the flow of operations in the Bayesian optimization are given in the Materials and Methods section. A description of the Bayesian implementation in Matlab for pulse sequence optimization, along with an explanation for constraining the pulse optimization, is provided in the Supplementary Information  Text and Figs. S1 and S2.

It is relevant to address the subtle difference between non-linear optimization and Bayesian-type artificial intelligence when optimizing multivariate functions with complicated impact on the DNP transfer performance. In this case, the observeable is the intensity accumulated in large spin systems with contributions coming from direct and many indirect (spin diffusion type) polarization transfer pathways. Although the objective may be the same, the learning process is completely different. Non-linear optimization using gradient or simplex methods\cite{press2007numerical,simplex} have no historical intrinsic learning (and thereby no predictive model) of the spin system response beyond the previous step, while the Bayesian approach accumulates learning from all previous steps. This may be a significant advantage in the optimization of pulse sequences, particularly those addressing unknown spin systems, influenced by instrument features in several transfer steps. This applies in particular when the set of observables for the iterative evaluation is small.

\subsection{Optimization of the DNP transfer pulse sequence for trityl}

\begin{figure*}[ht!]
    \centering
    \includegraphics[width=\textwidth]{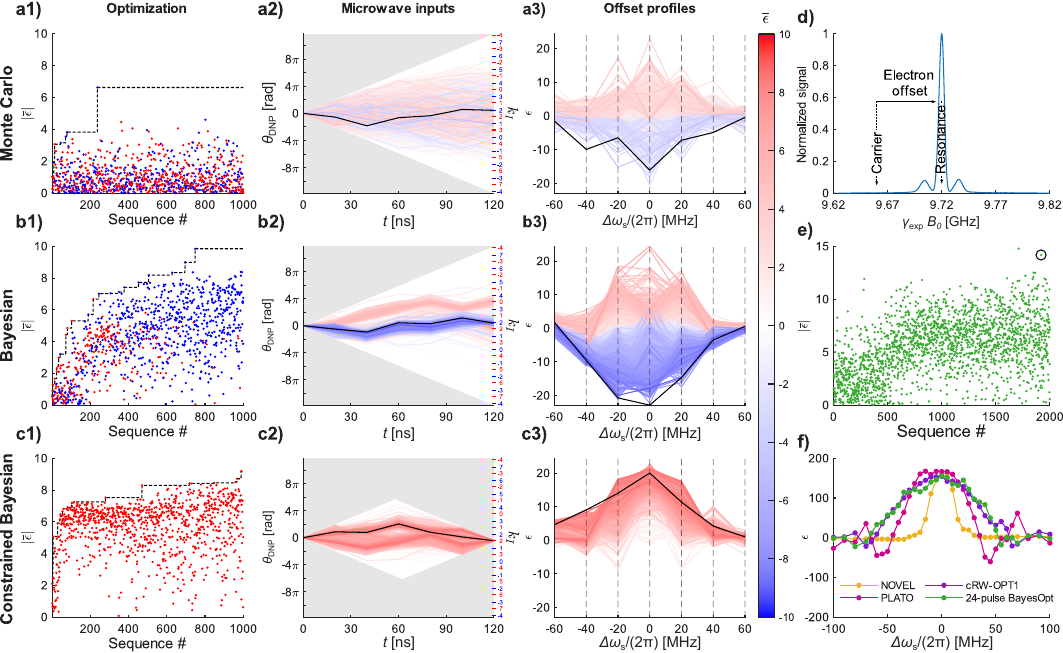}
    \caption{\textit{in situ} optimization of the DNP transfer element in Fig. \ref{fig:Setup}a with 6 (a-c)  and 24 (e,f)   pulses (modulation time $t_m$=120 ns; 49 MHz maximal MW field strength; initial flip pulse of amplitude 47 MHz and duration 5 ns, see Fig. \ref{fig:Setup}a). The measurements were performed at 80 K using a  trityl OX063 in glycerol-d$_5$ (a-e) and glycerol-d$_8$ (f). The DNP transfer element was repeated 1-30 (a-c) and 6 (e-f) times. The buildup time was 2 (a-c,e) and 64 (f) s. (a-c) Results for 3  optimization methods: a) Monte Carlo, b)  Bayesian, and c) constrained Bayesian (1000 evaluations each, 51 h optimization). (a1-c1) Absolute average enhancement $|\overline{\epsilon}|$  for the offset profile; red/blue dots represent match to ZQ/DQ resonances. (a2-c2) Accumulated nutation angle during each of the evaluated sequences. Colored numbers reflects $k_I$ values for ZQ (red) and DQ (blue) resonances according to Eq. (\ref{Eq1}). (a3-c3) Electron-spin offset profile. In columns 2) and 3) the color is weighted by the absolute average enhancement. The best sequences are marked in black. d) Different electron-spin offsets were realized by placing the MW carrier at different frequencies with respect to the  EPR resonance of trityl. e)   Constrained Bayesian optimization for a 24-pulse DNP transfer element (black circle marks the sequence with  best offset profile when repeated with higher resolution in electron-spin offset). f) Offset profiles for the 24-pulse BayesOpt sequence marked in e) compared to NOVEL, PLATO, and cRW-OPT1. Offset profiles for trityl in glycerol-d$_5$  can be found in Supplementary Information 
    Fig. S3. Settings for Bayesian optimizations are provided in Supplementary Information Tables 
    S1 and S2.}
    \label{fig:Figure2}
\end{figure*}

As a first example, demonstrating the relative performance of \textit{in situ} Bayesian optimization of pulsed DNP experiments, we optimize the DNP transfer element (using a prior fixed strong 90$^\circ$ excitation pulse) for a sample of 5 mM  trityl OX063
 in a H$_2$O:D$_2$O:Glycerol-d$_5$ solution (1:3:6 by volume) at 80 K. Trityl radicals are well characterized in terms of  electron-nuclear spin interactions, and they have been used extensively\cite{Mathies:2016aa} as probe for the performance of pulsed DNP experiments. This applies to experiments spanning from the simple NOVEL\cite{novel,offNovel} sequence to the more recent  advanced PLATO\cite{PLATO_adv} and cRW-OPT\cite{rCW_EEHT_DNP} methods. This facilitates direct comparison to state-of-the-art pulse sequences, which, for this well-known system, may be designed analytically and numerically with a remarkably good match to experimental data \cite{BEAM,PLATO_adv,rCW_EEHT_DNP,carvalho2026,offNovel,TOP_DNP,XiX_DNP,TPPM_DNP,DNP_steady_state}. An important aspect is that trityl, through the relatively narrow EPR resonance (see Fig. \ref{fig:Figure2}d), enables direct optimization and evaluation of pulse sequences in terms of electron spin resonance offset.

To demonstrate \textit{in situ} DNP pulse sequence development, Figure \ref{fig:Figure2} illustrates different aspects of optimizations with different  protocols  using the trityl radical system described above. Our aim is broadband DNP, which for the narrow-resonance (see Fig. \ref{fig:Figure2}d) trityl sample is invoked by adding the response from experiments executed with electron-spin resonance at different offsets; specifically, 0, $\pm$20, $\pm$40, and $\pm$60 MHz. For the optimization, we explore a DNP transfer element with  6 $x$-phase pulses each of 20 ns duration with the number repetitions of this element being a variable (between 1 and 30) in the optimization  and limiting the MW nutation frequency to a maximum of 49 MHz. The three explored optimization protocols are Monte Carlo (i.e., random MW amplitudes), Bayesian optimization, and Constrained Bayesian as illustrated in panels a, b, and c, respectively. In the latter case, we use constrained random walk to limit evaluations to cases fulfilling a DNP resonance. Using this setup, the columns in Figure \ref{fig:Figure2} illustrate, for each method, the outcome of  an optimization with 1000 evaluations (i.e., different MW pulse sequences). 
The left column (panels 1) displays the absolute value of the DNP enhancement ($\epsilon_{DNP}$; enhancement relative to direct excitation of the $^1$H spins) averaged over the specified electron-spin offsets for the 1000  pulse sequences with the dashed line accumulating the highest transfer efficiency. The second column (panels 2) shows the accumulated nutation angle  during the $t_m$ = 120 ns  pulse sequence element with one trajectory for each of the 1000 pulse sequence elements, where the electron-spin offset is set to zero. The accumulated nutation angle is defined as $\theta^{DNP} = \sum_{i=1}^6 \omega_i^{MW} \Delta t$ with $\omega_i^{MW}$ being the amplitude of the $x$-phase pulse $i$, the amplitude can be positive and negative, and $\Delta t$ = 20 ns the length of each pulse. The ticks and numbers in red and blue to the right of the trajectories mark the angles corresponding to a zero-quantum (ZQ, red) and double-quantum (DQ, blue) resonance using the definition (in angular units)
\begin{equation}
    \theta^{DNP}=\pm (\omega_{0I}t_m-k_I 2\pi) \quad .
    \label{Eq1}
\end{equation}
Here the $+$ and $-$ signs correspond to ZQ and DQ, respectively, $\omega_{0I}$ denotes the Larmor frequency of the protons spins, $\omega_m=2\pi/t_m$ is the modulation frequency of the MW pulse sequence element, and $k_I$ is an integer (the colored numbers correspond to different $k_I$ values).\cite{rCW_EEHT_DNP} A more detailed description of these parameters and the resonance condition is given in the Materials and Methods section.
The third column (panels 3) shows the detected 
enhancements $\epsilon$ for the 1000 pulse sequences as a function of electron spin resonance offset with the dashed lines indicating the offsets used in the optimization. We note that sequences matching or approaching a ZQ resonance have positive intensity (marked in red in correspondence with the angle trajectories in proximity of a ZQ resonance) and, likewise, sequences matching  or approaching a DQ resonance are associated with negative intensity (marked in blue).

Addressing specifically the Monte Carlo (i.e., random pulse amplitudes within the constraints of the maximum pulse amplitude) optimization in Fig. \ref{fig:Figure2}a, it appears that among 1000  sequences a relatively equal mixture of pulse sequences in proximity of ZQ or DQ resonances over at very broad span of $k_I$ values is found. Many of those, however, do not match perfectly, resulting in destructive ZQ/DQ overlap or low transfer due to MW inhomogeneity (\textit{vide infra}). This is evidenced by the enhancement factors in Fig. \ref{fig:Figure2} a1, with the maximum approaching 7, but the average/standard deviation (SD) being 1.0/0.87, and is also seen from the offset profiles in Fig. \ref{fig:Figure2} a3. 

Significantly better results are, as demonstrated in Figure \ref{fig:Figure2}b, obtained by optimization of 1000 pulse sequences using machine-learning Bayesian optimization of  feedback-loop intensities from  DNP experiments. It is clear that the sequences rapidly improve through the Bayesian optimization, reaching a maximum enhancement factor of 10 and average/SD values of 4.1/2.1 for all sequences and 5.4/1.8 for the last 200 evaluations. It is seen here that the sequences induced by optimization concentrate around the two $k_I$ values, 0 (ZQ) and 2 (DQ) (see Fig. \ref{fig:Figure2} b2), and provide more intense/broadband profiles (see Fig. \ref{fig:Figure2} b3). It is evident that Bayesian optimization results from a learning process with many of the sequences being trials, and thereby likely associated with low enhancement (see Fig. \ref{fig:Figure2} b1), while it searches for the best sequences. The best sequences are represented by the points with highest integrated enhancement factors. In this particular case, a relatively high fraction of the best pulse sequences is also of DQ nature.

It is observed that the Bayesian optimizations find the best sequences for particular values of $k_I$. This finding can be related to our previously established knowledge that DNP pulse sequences with finite, but not too large  effective fields works best as they are not excessively influenced by MW inhomogeneity. \cite{PLATO_adv,rCW_EEHT_DNP} This motivates the combination of Bayesian optimization with constraints towards low values of $k_I$. This may be achieved by constraining the search space of the Bayesian optimization by constrained random walk (cRW) procedures as described earlier\cite{rCW_EEHT_DNP} in combination with EEHT\cite{EEHT} and non-linear optimization. An example of the performance of such constrained Bayesian optimization, with the restrictions of all pulse sequences to fulfill the ZQ resonance condition $k_I$ = 2, is illustrated in Fig. \ref{fig:Figure2}c. The enhancement factor plot (Fig. \ref{fig:Figure2} c1) shows exclusively ZQ behavior and a much faster convergence to high transfer efficiencies. The effect of constraining the optimization to the $k_I$=2 resonance condition is clearly visible in Fig. \ref{fig:Figure2} c2 with the result of broadband sequences with high enhancement as seen in Fig. \ref{fig:Figure2} c3. We should note that, upon close inspection, the Bayesian and constrained Bayesian optimizations in this particular case provide the best sequence in the non-restricted Bayesian optimization with the accumulated angle $\theta^{DNP}$ being around 23$^\circ$ below the $k_I$ = 2 DQ resonance at 81$^\circ$. This deviation may be ascribed to effects from electron-spin offset and MW inhomogeneity. It is emphasized that this particular sequence has the lowest effective field, indirectly reinforcing our statement that large effective fields render the pulse sequences more susceptible to MW inhomogeneity, which in this practical setup is significant.\cite{PLATO_adv,rCW_EEHT_DNP}. 

To further document the important finding that \textit{in situ} optimization finds sequences least susceptible to MW inhomogeneity effects, Supplementary Information Fig. 
S4 maps the enhancement factor as a function of $\theta^{DNP}$ for all optimizations in Fig. \ref{fig:Figure2}. It is clearly evident that the best sequences are associated with the ZQ $k_I$ = 2 resonance characterized by a linear field of $\theta^{DNP}/t_m= 1.87$ MHz. This happens with less probability for the Monte Carlo optimization but occurs clearly for the unconstrained Bayesian optimization. The latter provides two bands with the largest enhancement seen for the DQ $k_I=2$ resonance and around half enhancement for the ZQ $k_I=0$ resonance, with a linear field of $\theta^{DNP}/t_m= 14.8$ MHz being identical to the less broadband NOVEL experiment. We should also direct attention to Supplementary Information Fig. S5,
illustrating the distribution of sequences in Fig. \ref{fig:Figure2} and their enhancements as a function of the number of repetitions of the building block.  All of them provide good sequences except the 1 - 4 repetitions with the overall transfer element being too short for the present pseudo-secular hyperfine coupling to ensure good polarization transfer.

For further testing of the Bayesian optimization approach, the experimental optimizations presented in Figs. \ref{fig:Figure2} a1-c3 were also investigated with an \textit{in silico} approach. Here, the evaluation of a given pulse sequence was not based on the experimental transfer performance but rather on the performance of a simulated polarization transfer for a two-spin e$^-$-$^1$H
system. Additional information on numerical simulations is presented in the Materials and Methods section. The numerical analysis is presented in Supplementary Information Fig. 
S6 with corresponding optimization settings in Table 
S3. By comparison of the \textit{in silico} and \textit{in situ} approaches, it is seen that the features are similar but with the trend that the experimental results provide more broadband performances which. This can, to a certain extent, be ascribed to the line width of the EPR line (Fig. \ref{fig:Figure2}d) which for the numerical calculations is assumed infinitely narrow.

With these comparative performance evaluations at hand, we proceeded with  constrained Bayesian optimization of pulse sequences with more pulses. \textit{In silico} and \textit{in situ} constrained Bayesian optimization employing 6, 12, and 24 pulses for the transfer element are presented in Supplementary Information Figs. 
S7 and 
S8 with corresponding optimization settings in Tables 
S4 and 
S5, respectively. From these data, it is evident that by increasing the number of pulses, the performance of the transfer elements in terms of bandwidth does get better. However, the cost of increasing the number of variables/pulses for the Bayesian optimization is that it takes more evaluations to find good hits.  We direct attention  to   optimization of 24-pulse DNP elements as illustrated in Fig. \ref{fig:Figure2}e, leading to an enhancement factor of 15 (averaged over the selected offset points). The broadband excitation profile of this 24-pulse DNP pulse sequence (pulse sequence, henceforth denoted BayesOpt, is  given in Supplementary Information Table 
S6) was then compared to NOVEL, PLATO, and cRW-OPT1 pulse sequences in Fig. \ref{fig:Figure2}f. The sequences provide enhancement factors approaching 180 and the average enhancements, when integrated over the offset range from -60 MHz to 60 MHz, of 98 (cRW-OPT1), 35 (NOVEL), 69 (PLATO), and 96 (BayesOpt), respectively, for a trityl OX063 glycerol-d$_8$ sample (64 s buildup). We note that by using the Bayesian \textit{in situ} experiment design protocol, we found pulse sequences with similar broadband excitation profile as the previously state-of-the-art analytical-numerical designed cRW-OPT1 sequence.\cite{rCW_EEHT_DNP}  This demonstrates that, on this  well-characterized  trityl OX063 system, \textit{in situ} Bayesian optimization enables design of pulse sequences with comparable bandwidth to those designed with the best methods developed so far. This validates the usefulness of \textit{in situ} Bayesian experimental optimization, providing a strong case for optimization on systems without similar knowledge of the involved spin system. This may lead to superior pulse sequences and,  importantly, increase spin engineering insight into the function of DNP experiments in complex spin systems and/or non-ideal instrumental circumstances.

\subsection{Optimization of the excitation pulse of the DNP experiment for trityl}
\begin{figure*}[ht!]
    \centering
    \includegraphics[width=\textwidth]{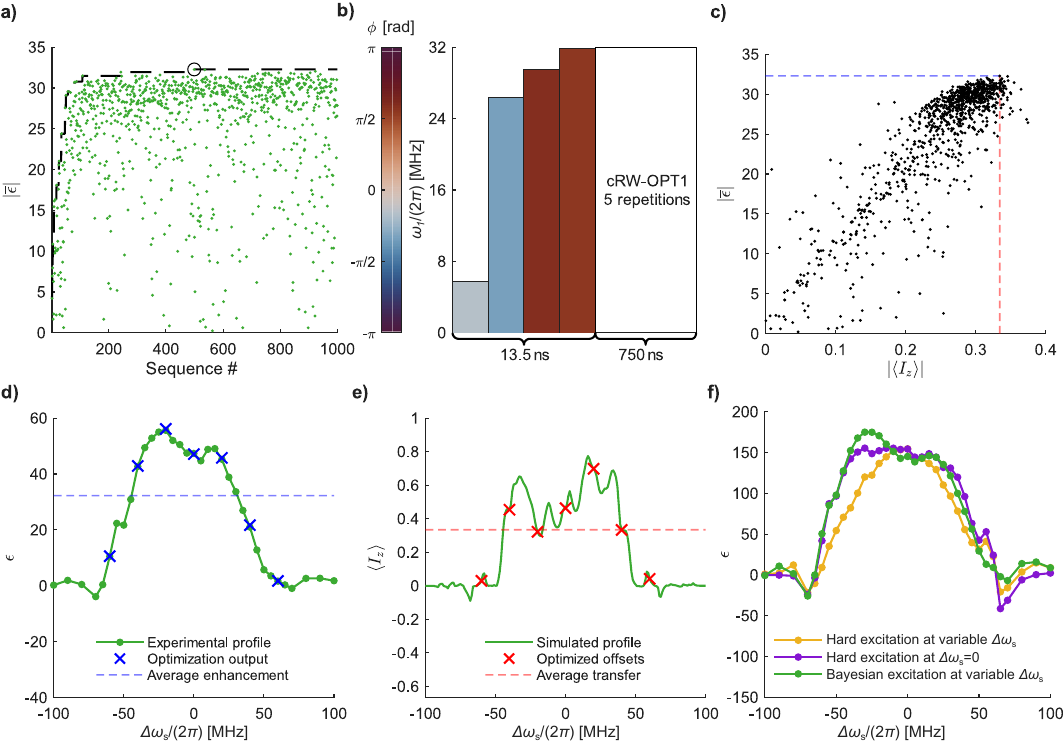}
    \caption{a) Bayesian \textit{in situ}  optimization of a 4-pulse excitation sequence with flexible total duration of 5-20 ns (all pulses equal length, amplitude up to 45 MHz,  free phase) followed by cRW-OPT1 DNP transfer (5 repetitions; 5 s buildup) using  trityl OX063 in glycerol-d$_5$.  b) The best 4-pulse excitation sequence (marked by circle in a); Supplementary Information Table 
    S7) followed by cRW-OPT1 DNP transfer.  c) Comparison of simulated (horizontal) and experimental (vertical) enhancements for all sequences in a). c) Offset profile for the best excitation sequence. Blue crosses show the enhancements found during optimization, green dots show repetition of the experiment for a finer grid of offsets, and the dashed line shows the average. e) Simulated profile of the best excitation sequence. Crosses show simulated transfer at the optimized offsets, and the dashed line shows the  average. Note that the dashed lines in c) correspond to the best sequence as shown in d) and e)
    f) Offset profiles for cRW-OPT1 using the best excitation sequence and a hard excitation pulse (amplitude 32 MHz). The carrier of the hard excitation pulse was either following the  offset for the DNP transfer sequence or fixed at the trityl resonance. These experiments used trityl OX063 in glycerol-d$_8$ and a buildup time of 64 s. Settings for the Bayesian optimizations are given in Supplementary Information Table 
    S8.}
    \label{fig:FlipOpt}
\end{figure*}

The most common\cite{Pulsepol,XiX_DNP,TPPM_DNP,carvalho2026} DNP transfer elements mediate transverse spin-locked electron-spin $S_x$ to proton-spin $I_z$ polarization transfer. This implies that a typically short and intense excitation pulse is required to generate transverse $x$-phase electron-spin coherence prior to polarization transfer via spin locking. Recently, in the development of highly broadband DNP experiments, such as the cRW-OPT pulses sequences,\cite{rCW_EEHT_DNP} we demonstrated that the initial pulse may impede the broadband capability of the overall experiment. As such, for transverse spin-lock DNP, it is relevant to address optimization of the initial excitation pulse as well as the DNP transfer element. To demonstrate diversity in the use of \textit{in situ} experimental Bayesian pulse sequence optimization, Fig. \ref{fig:FlipOpt} illustrates optimization of the phases and amplitudes of an initial 4-pulse excitation sequence. The optimization allowed flexibility of the total duration to be between 5 and 20 ns to optimally match DNP transfer with a previously developed 30-pulse (150 ns total, 5 ns pulses) cRW-OPT1 broadband DNP transfer element.\cite{rCW_EEHT_DNP} This implies that the optimization considers both excitation and transfer while only optimizing the former element. The optimized target pulse element is not necessarily a perfect broadband $\pi/2$ pulse but a pulse that works optimally with the specific cRW-OPT1 element in generating the best transfer from electron to nuclear spin polarization. In this case, the Bayesian optimization was unconstrained.

Figure \ref{fig:FlipOpt}a reports on 1000 evaluations of an \textit{in situ} Bayesian optimization of a 4-pulse excitation pulse, with green dots representing the transfer of each sequence and the dashed black line representing the fast establishment of a large DNP enhancement factor for the combined excitation pulse and cRW-OPT1 pulse sequence. The pulse amplitudes could reach a maximum of 45 MHz, the phase of each pulse was allowed to vary freely, and the total length of the excitation pulse sequence could vary between 5 and 20 ns. The optimal excitation pulse sequence, represented by the black circle, is illustrated in Fig. \ref{fig:FlipOpt}b concatenated with the cRW-OPT1 pulse sequence (repeated 5 times). In Fig. \ref{fig:FlipOpt}c, we investigate the correlation between average enhancement factors obtained experimentally (vertical) and numerically (horizontal) for the 1000 excitation pulse sequences evaluated in Fig. \ref{fig:FlipOpt}a. A perfect match between simulations and experiments would yield a straight-line 
correlation, and would indicate that the two-spin system indeed does represent part of the spin system. Deviation from a linear correlation, as seen, illustrates that the \textit{in situ} optimization captures either spin-system or instrumental aspects that are not included in the theoretical/numerical model used for the simulations.  

Figure \ref{fig:FlipOpt}d illustrates the experimental broadband profile of the combined optimized excitation pulse sequence and cRW-OPT1 DNP transfer pulse sequence. The pulse sequence is evaluated at the offset points used in the optimization marked by blue crosses and repetition of the experiments with a finer grid of offset values represented by the green dots and line. This demonstrates that optimization with relatively few points covers well the full span of electron-spin offsets. This may be naturally ascribed to the 10-15 MHz linewidth of the trityl radical, the effect of which may be rationalized by folding of the offset profile with the shape of the trityl EPR line (see Fig. \ref{fig:Figure2}d). The offset profiles in Fig. \ref{fig:FlipOpt}e represent numerical simulations of the optimal pulse sequence calculated for the offsets used in the optimization as well as for a finer set of offset points. The two-spin simulations indeed validate the broadband behavior identified experimentally; however, with more pronounced intensity variations, which may be ascribed to insufficient modeling of the true spin dynamics by an electron-nuclear spin-pair system, as already hinted at by Fig. \ref{fig:FlipOpt}c. Figure  \ref{fig:FlipOpt}f compares cRW-OPT1 measurements conducted using either the Bayesian optimized excitation pulse or single-pulse (amplitude 45 MHz, length 5 ns) excitation. The single-pulse excitation was executed in two ways: With the carrier frequency of the excitation pulse on-resonance (i.e., $\Delta \omega_S/(2\pi)$ = 0) or, experimentally more relevant, with the carrier moving with the offset of the transfer element (i.e., $\Delta \omega_S/(2\pi)$). The latter of which was also used for the Bayesian 4-pulse sequence. It is evident that the Bayesian excitation pulse effectively prevents losses at the outer parts of the offset profile as achieved by the  single-pulse  matching the “idealized” case, where the excitation pulse remains always on-resonance (i.e., $\Delta \omega_S/(2\pi)$ = 0). This is obviously important for implementation for systems with broad EPR lines.

\subsection{Optimization on TEMPO}

\begin{figure*}[ht!]
    \centering
    \includegraphics[width=\textwidth]{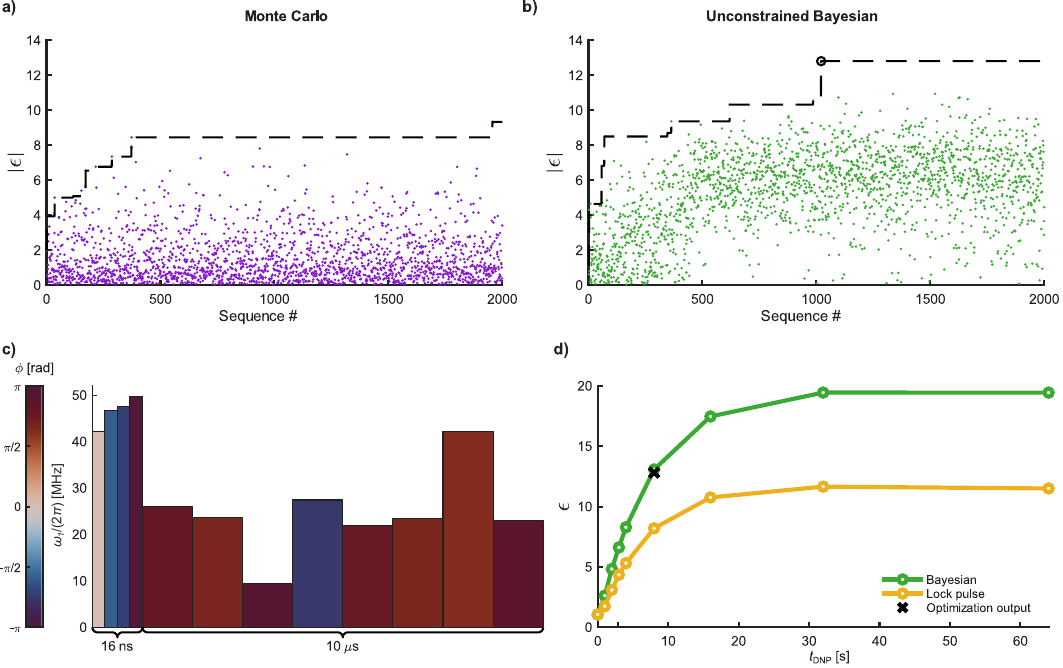}
    \caption{\textit{In situ} experimental Monte Carlo and Bayesian optimization of excitation pulse and polarization transfer elements of pulsed DNP experiments for 2 mM sample of 4-Oxo-TEMPO. a) Evaluated enhancements for Monte Carlo optimization of  control variables. b) Evaluated enhancements for an unconstrained Bayesian optimization of  control variables. c) The best pulse sequence is marked by the circle in b). The duration of the excitation pulses was 4 ns each, and transfer pulses were 1250 ns each (the time axis is not to scale). The pulse sequence is given in Supplementary Information Table 
    S11. d) Buildup curve for the best sequence found using Bayesian optimization compared to the buildup for a hard excitation pulse of 52 MHz for 5.5 ns followed by a spin-lock pulse of constant amplitude of 31.2 MHz for 10 $\mu s$. The cross showed the enhancement measured during the optimization with a buildup time of 8 s. Settings for the Bayesian optimizations are given in Supplementary Information Table 
    S12.}
    \label{fig:Tempo}
\end{figure*}

While the narrow EPR line of trityl and its well-studied spin dynamics\cite{BEAM,PLATO_adv,rCW_EEHT_DNP,carvalho2026,offNovel,TOP_DNP,XiX_DNP,TPPM_DNP,DNP_steady_state} make this sample ideal for investigating broadband DNP transfer, it  simultaneously makes it a less interesting use-case for DNP transfer in more typical EPR cases that are influenced by larger hyperfine couplings, electron-spin coupling to different nuclear spin species, and broad EPR resonances.  Therefore, to explore the use of learning-based \textit{in situ} experimental design of DNP pulse sequences for a more challenging system, where current analytical-numerical design strategies may not yet suffice, Figure \ref{fig:Tempo} illustrates Bayesian optimization of DNP for the 2 mM nitroxide-radical 4-Oxo-TEMPO in a DMSO-d$_6$:D$_2$O:H$_2$O solution (6:3:1 by volume) at 80 K.
In this case, the unpaired electron is strongly coupled to $^{14}$N (spin $I$=1) and further to numerous $^1$H spins implying that it is not straight-forward to assume the same resonance conditions as used in the constrained Bayesian approach demonstrated above. 

Initial \textit{in situ} tests were performed to determine the overall structure of the pulse sequence to be optimized, e.g., trial setups to determine if initial flip pulses were desirable or the number of pulses to be used for efficient Bayesian optimizations. These exploratory data are presented in Supplementary Information Fig. 
S9 and corresponding Bayesian settings in Tables 
S9 and S10. On the basis of these results, DNP pulse sequences involving both the excitation pulse and the subsequent DNP transfer element were optimized using unconstrained Bayesian optimization and compared with Monte Carlo (random) optimization. The excitation pulse involved 4 pulses with a maximum amplitude of 49 MHz, free phase, and a total length of the excitation pulse sequence of 5-20 ns, while the DNP pulse sequence element comprised 8 equally long pulses spanning 10 $\mu$s with free phase variation and maximum amplitude of 52 MHz. We note that this transfer element is not repeated, and thereby the sequence representing the testing object in this case is very different from the ns-scale pulse sequences used above for the case of trityl. We also note that this aspect, and with this which method will be best in general terms, will be subject to further exploration. It is, however, evident from Fig. \ref{fig:Tempo} that  Bayesian optimization (Fig. \ref{fig:Tempo}b) provides better sequences (enhancement factor of 13) than those resulting from a similar 2000 sequence Monte Carlo optimization (Fig. \ref{fig:Tempo}b, enhancement factor of 9). 

The best Bayesian optimized pulse sequence (corresponding to the circle in Fig. \ref{fig:Tempo}b) is shown in Fig. \ref{fig:Tempo}d. Comparison of this sequence with a spin-lock pulse sequence with MW amplitude 31 MHz is illustrated in Fig. \ref{fig:Tempo}d revealing a gain in sensitivity by 70 \% when using the Bayesian-optimized DNP pulse sequence. Three points should be noted here: ($i$) the spin-lock sequence used a larger amplitude than NOVEL. However, as seen in Supplementary Information Fig. 
S10, quite similar efficiency is observed at the true NOVEL condition, ($ii$)  the broadband DNP pulse sequences did not work better than the spin-lock sequence in this case, and ($iii$)  for the purpose of illustration, we deliberately chose a low concentration of TEMPO to focus on solid-effect DNP rather than the higher concentration typically generating DNP by the cross effect. These aspects will independently be subject to further investigation.

 \section{Discussion}
 \label{sec:Discussion}

We have in this work explored the use of learning-based optimization methods to develop magnetic resonance pulse sequences \textit{in situ} on a spectrometer. Our target has been DNP utilizing Bayesian-based learning methods in a feedback-loop control system using a home-built X-band pulsed DNP/EPR spectrometer with fast AWG's and strong MW amplification. This setup was mimicked by a corresponding numerical optimization protocol based on density operator calculations. 

The first target was trityl OX063 in a “DNP juice” solution, which represents a widely used system for method development and applications of DNP. Various pulsed DNP experiments have recently been developed and characterized for trityl, which here represents a “model system” to explore Bayesian-based learning procedures for \textit{in situ} method development. For this system, we could identify: ($i$) Monte Carlo-based \textit{in situ} pulse sequence optimization is capable of producing good pulse sequences, but ($ii$) the association with Bayesian learning procedures  provides better sequences faster, and ($iii$) both faster convergence and overall better performing sequences can be obtained by limiting the search space for the Bayesian learning by constrained random walk procedures. The latter operate by restricting the prediction-model-based search to sequences fulfilling ZQ or DQ resonances. Remarkably good pulse sequences were found, with similar performance to the best sequences developed for this well-characterized system by analytical and numerical methods. The Bayesian approach clearly found sequences that were broadband in terms of electron spin offset and efficiently compensated for experimental MW inhomogeneity. Furthermore, it also identified that the pulse sequence performance can be improved beyond the limits imposed by a simple two-spin system. This is supported by deviations from a linear correlation between experimental data and numerical calculations for the experimentally developed pulse sequences. Overall, the experiments on the trityl system may be considered a first proof of concept for \textit{in situ} learning-based pulse sequence optimization.

Taking a more challenging system where solid-effect-type pulse sequences may be less efficient, as here represented by the low-concentration (2 mM) sample of the 4-Oxo-TEMPO nitroxide, we demonstrated  improvements of the DNP enhancement by 70 \% relative to the NOVEL experiment. The overall transfer efficiencies are not impressive, but these first experiments serve as a compelling case for the use of \textit{in situ} learning-based methods to further develop pulsed DNP experiments and for potential inspiration to analytical-numerical design protocols. The proposed \textit{in situ} pulse sequence design may find applications in wider ranges of magnetic resonance, including liquid- and solid-state NMR,  EPR, and the DNP approach demonstrated here may be of considerable interest for electron-spin-based quantum sensing and information technology.

\section{Materials and Methods}
\label{sec:Methods}

\subsection{Experimental Setup}
\textit{In situ} experimental optimization and evaluation of pulsed DNP experiments were carried out on  home-built X-band pulsed EPR/DNP spectrometer (based on the design of Doll \textit{et al.} \cite{Doll:2017wb}) equipped with a Keysight M8190A 14 bit/8GSa and Z\"urich Instruments HDWAG4 16 bit/2.4 GSa (Z\"urich Instruments, Z\"urich, CH) arbitrary waveform generators on electron and nuclear spins, respectively, a 2 kW Applied Systems Engineering 176 X TWT MW amplifier, a SpinCore  GX-5 iSpin NMR console (SpinCore Technologies Inc., Gainesville, FL), a 300 W NanoNord RF amplifier (NanoNord A/S, Aalborg, Denmark), and a Bruker MD4 electron-nuclear double resonance probe (Bruker BioSpin, Rheinstetten, DE) extended with an external tuning and matching circuits.

\begin{figure*}[ht!]
    \centering
    \includegraphics[width=\textwidth]{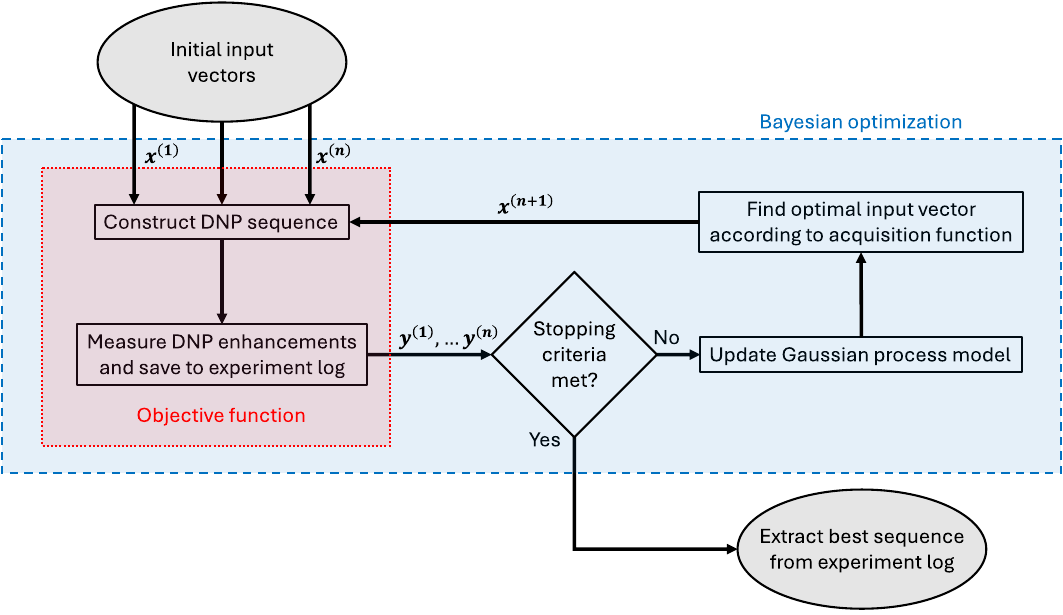}
    \caption{Diagram showing the procedure used for {\it situ} Bayesian learning-based optimization. First, a set of random input vectors is given to the objective function. Each of these vectors is then used to construct a pulsed DNP sequence, which is executed on the experimental setup. Each sequence and the resulting DNP enhancements are stored in an experiment log for future reference. After the initial input vectors have been evaluated, a Gaussian process model is initialized from the experiment log, and the chosen type of acquisition function is optimized based on this model. The (approximate) optimizer of the acquisition function is then used as the input vector for the next evaluation of the objective function, completing the feedback loop. When certain user-defined conditions, such as number of iterations, are met, the loop is set to stop, and, subsequently, the best sequence can be extracted from the experiment log.}
    \label{fig:flowchart}
\end{figure*}

All experiments used the pulse sequence in Fig. \ref{fig:Setup}a, basically involving an MW channel for polarization transfer and an RF channel for proton detection. The pulse sequence initially saturates $^1$H with a set of $S=11$ pulses of duration 1.32 $\mu$s separated by $\tau_{\text{sat}}=1$ ms. The $^1$H signal was detected using a solid-echo sequence  $\pi / 2 - \tau - \pi / 2$ with $\tau_\text{SE}$ = 25 $\mu$s;  RF pulses typically used an RF field strength of 227.3 kHz, corresponding to a $\pi/2$ pulse time of 1.10 $\mu$s.  Each experiment used a repetition time of $\tau_{\text{rep}}=2$ ms. The pulse sequence and signal detection were executed as part of the optimization flow diagram shown in Fig. \ref{fig:flowchart}. Experimental polarization enhancements (denoted $\epsilon_{\text{p}}$) are defined by the ratio between the DNP-enhanced signal intensity and the thermal equilibrium signal intensity.
For each experimental session, a series of calibrations was done  to determine the MW resonator profile (maximum available MW field strength), with a representative example given in Supplementary Information Fig. 
S11.

An EPR field sweep of the three samples is presented in Supplementary Information Fig. 
S12. Experimental evaluation $T_1$ relaxation times for electron spins and protons for 4-Oxo-TEMPO and  trityl OX063 in glycerol-d$_5$ samples are presented in Supplementary Information Fig. 
S13.   

\subsection{Samples}

Experiments were conducted at 80 K on samples of 5 mM  trityl OX063 in a H$_2$O:D$_2$O:Glycerol-d$_5$ solution (1:3:6 by volume; note the higher concentration of protons in this system relative to the commonly used DNP juice, which, due to its higher concentration of $^1$H, has lower overall polarization transfer), 5 mM  trityl OX063 in a H$_2$O:D$_2$O:Glycerol-d$_8$ solution (1:3:6 by volume; with a lower concentration of $^1$H),  and 2 mM 4-Oxo-TEMPO in DMSO-d$_6$:D$_2$O:H$_2$O solution (6:3:1 by volume). 

\subsection{Bayesian Optimization}
\label{sec:Bayes}

Pulsed DNP sequences were optimized \textit{in situ} on the spectrometer with a feedback-loop control shown in Fig. \ref{fig:flowchart} using Bayesian optimization, which was originally developed as a black-box optimization method in machine learning. Bayesian optimization is inspired by Bayesian statistical methods to determine a probabilistic surrogate model of the objective function \cite{Bayes,BayesDesign}. The objective feedback DNP enhancements $f(x)$ are obtained from a pulse sequence, which is in turn determined from an input vector $x$ of variables as described in Supplementary Information (Supplementary Information Text and Figs. 
S1 and 
S2). Given the underlying stochastic dynamics of the spin system and instrumentation, we model the observed feedback DNP enhancement for a given input vector $x$ stochastically as $y = f(x) + \varepsilon$ for independent Gaussian noise, $\varepsilon \sim N(0,\sigma^2)$, where the standard deviation $\sigma > 0$ summarizes the measurement error of the experimental setup. The Bayesian approach to this regression model now postulates to encode our uncertainty about the feedback model $f$ by treating it as a stochastic object. To this end, we first put a rather uninformative prior distribution on $f$ and then sequentially update our beliefs on its nature from obtained experimental evidence $y^{(i)} = f(x^{(i)}) + \varepsilon^{(i)}$, $i =1,\ldots,n$, for implemented input vectors $x^{(1)},\ldots,x^{(n)}$.  Specifically, we put a Gaussian process prior \cite{rasmussen} on the objective function $f$ such that before collecting experimental evidence, the feedback is modeled to be drawn from a multivariate normal distribution
\begin{equation}
    f(x^{(1)}), \dots ,f(x^{(n)}) \sim N( M,\Sigma) \quad ,
    \label{eq:SurModel} 
\end{equation}
 where $M$ is a mean vector with elements $M_i  =  \mu_0(x^{(i)})$ and $\Sigma$ is a covariance matrix  with elements $\Sigma_{ij}  =  \Sigma_0(x^{(i)},x^{(j)})$. The functions $\mu_0$ and $\Sigma_0$ are called the mean function and kernel, respectively, and are chosen prior to the optimization. For the observed feedback DNP enhancements $y^{(i)} = f(x^{(i)}) + \varepsilon^{(i)}$ on the spectrometer, we therefore obtain the Gaussian prior 
\[y^{(1)}, \dots ,y^{(n)} \sim N( M,\Sigma + \sigma^2\mathbb{1}) \quad . \]
Our implementation of Bayesian optimization used a built-in MatLab function \matlab{bayesopt} that uses a constant mean function and the ARD Matérn 5/2 kernel \cite{BayesKernel}. 

The optimization is initialized by a number of Monte Carlo evaluations; in our case, 4. The feedback is used to determine the hyperparameters used for the prior in Eq.\ (\ref{eq:SurModel}). We can then predict the feedback $f(x^{(n+1)})$ at a new input vector $x^{(n+1)}$ by calculating its posterior distribution given evaluations $y^{(1)}, \ldots, y^{(n)}$ and corresponding input vectors $x^{(1)},\ldots, x^{(n)}$ using Bayes' formula.  This conditional distribution is explicitly given by
\begin{equation}\label{eq:post}
\begin{split}
     f(x^{(n+1)})& \mid y^{(1)},\ldots,y^{(n)}, x^{(1)},\ldots, x^{(n)}  \\  \sim  N&\big(\mu_n(x^{(n+1)}),\sigma_n^2(x^{(n+1)})\big) \quad ,
\end{split}
\end{equation}
where the expectation and variance functions $\mu_n$ and $\sigma_n^2$ are  computable from $\mu_0$, $\Sigma_0$, $\sigma$, the input vectors $x^{(1)},\ldots,x^{(n)}$ and the corresponding noisy DNP enhancements $y^{(1)},\ldots,y^{(1)}$ \cite{rasmussen}. To determine which input vector to evaluate next, an acquisition function $\mathrm{Ac}(x^{(n+1)})$ is used as a way to specify which attributes are desired for the posterior in Eq. \eqref{eq:post}. Specifically, it is designed to balance an inherent exploration-exploitation tradeoff, thus chosen to be  increasing in the posterior expectation $\mu_n(x)$ to exploit the learned information, $f$, in the first $n$ steps while also including a penalty term involving the standard deviation, $\sigma_n(x)$, to promote further exploration in the search space. An (approximate) optimizer $x^{(n+1)}$ of the acquisition function is then determined and translated to a new pulse sequence in the next experiment. This procedure is applied recursively to select the next experiment based on all previous experiments and feedback until a chosen stopping criterion is met. In our case the criteria was a maximal number of evaluations, and after this the best DNP enhancement and its corresponding pulse sequence could be extracted from the experiment log. An \textit{in silico} constrained Bayesian optimization using four different acquisition functions is shown in Supplementary Information Fig. 
S14 with settings given in Tables 
S13 - S16. Based on this analysis, it was decided to use the expected-improvement-plus acquisition function further on in this work.  

\subsection{Resonances and Constrained Random Walk}
\label{sec:cRW}

As a gradient-free learning algorithm, Bayesian optimization will typically be implemented without any constraints to freely modulate the DNP contact pulse sequence. In the interest of optimization speed, it may, however, be useful to guide or constrain the optimizations if the system to be optimized can be associated with  variables that rationally can be used to constrain the search and learning process. In the case of the DNP transfer, as illustrated by the trajectories in Fig. \ref{fig:Figure2} b2, the best solutions may approach ZQ or DQ resonances, as defined in Eq.\ (\ref{Eq1}). While this is to be expected only in cases with dominant two-spin interactions, as may be the case for the  trityl OX063 system using relatively short DNP transfer times, it may be relevant to explore the potential of adding constraints to Bayesian optimization, to improve speed of convergence, while knowing that this spin-system-wise may restrict the search space. We apply, as an example, the constrained Random Walk procedure described recently in combination with EEHT-driven non-linear optimization of DNP pulse sequences.\cite{rCW_EEHT_DNP}  This involves constraining the search to MW pulse sequences whose accumulated nutation angle $\theta^{DNP}$ fulfills Eq. (\ref{Eq1}). This  condition relates to the spin dynamics of the DNP experiment as described in the following.

In the simple case of a hyperfine-coupled electron (S)-nuclear (I) spin-pair system the Hamiltonian may, in the electron-spin rotating frame, be expressed as
\begin{equation}
    \mathcal{H} = \omega_{0I} I_z+\Delta\omega_S S_z+S_z (A I_z+B I_x)+\mathcal{H}_{MW}(t),
    \label{eq:2spins}
\end{equation}
where $\omega_{0I}$ is the nuclear Larmor frequency, $\Delta\omega_S = \omega_S-\omega_{MW}^{carrier}$ is the offset of the electron spin relative to the MW carrier at frequency $\omega_{MW}^{carrier}$, $A$ is the secular hyperfine coupling, $B$ is the pseudosecular hyperfine coupling, and $\mathcal{H}_{MW}(t)$ is the Hamiltonian of the MW irradiation in the rotating frame. 

In the cyclic interaction frame of both the electron and nuclear spins, it is possible, by single-spin-vector effective Hamiltonian theory (SSV-EEHT),\cite{SVEHT_EEHT} for a pulse sequence element with modulation time $t_m$ to recast the Hamiltonian as\cite{BEAM,PLATO_adv,DNP_echo,DNPbridge}
\begin{equation}
\begin{split}
    \widetilde {\mathcal{H}} = & \sum_{\kappa=x,y,z} \sum_{k_S=-\infty}^\infty a_{\kappa z}^{(k_S)}e^{ik_S\omega_m t}\widetilde S_\kappa (A I_z \\ & + \frac{B}{2}(e^{ik_I\omega_mt}I^++e^{-ik_I\omega_mt}I^-))  \\ &\quad- \omega_{eff}^{(S)}\widetilde S_z + \omega_{eff}^{(I)} I_z \quad ,
    \end{split}
    \label{eq:intframeH}
\end{equation}
where $\omega_m = 2\pi/t_m$ is the angular modulation frequency of the pulse sequence, $k_I$ is an integer, $\widetilde S_\kappa$ is the effective-field-frame electron-spin operators, $I_z$, $I^+$, and $I^-$ are nuclear-spin operators,  and $a_{\kappa z}^{(k_S)}$ are Fourier coefficients induced by the applied frame transformations. The two last terms describe the effective field for the electron and nuclear spins. 

From Eq. (\ref{eq:intframeH}), it is evident that matching of values $k_S=\pm k_I$ may lead to time-independent Hamiltonian components. Taking the first-order average Hamiltonian, it becomes evident that  resonances occur when the nuclear-spin effective field $\omega_{eff}^{(I)}=\omega_{0I}-k_I \omega_m$ matches the electron-spin effective field, $\omega_{eff}^{(S)}$, which lead to the resonance condition 
\begin{equation}
    \omega_{0I} - k_I \omega_m = \mp \omega_{eff}^{(S)}  \quad ,
    \label{eq:ssvres}
\end{equation}
with $k_I=\text{round}(\omega_{0I}/\omega_m)$. The negative and positive signs correspond to zero-quantum (ZQ) and double-quantum (DQ) resonances, respectively, leading to $\widetilde S_z \rightarrow I_z$ (ZQ) or $\widetilde S_z \rightarrow -I_z$ (DQ) polarization transfer. 

The condition in Eq. (\ref{eq:ssvres}) implies that ZQ or DQ resonances, and thereby DNP transfer, are obtained whenever the accumulated MW nutation angle $\theta^{DNP}$ matches the nutation angle of the electron spin $\omega_{0I}t_m-k_I 2\pi$ as outlined in Eq. (\ref{Eq1}). This feature can be used to constrain the space of solutions for the Bayesian optimization by simply requiring that the pulse sequences in the optimization fulfill this condition, as described in more detail in Supplementary Information Text. We note that the first-order effective Hamiltonian through the Fourier coefficients also provides information about the scaling factors for the ZQ and DQ Hamiltonians and thereby the time needed for efficient polarization transfer as described elsewhere.\cite{BEAM,PLATO_adv,DNP_echo,DNPbridge}

\subsection{Numerical simulations}
\label{sec:Num}

Numerical simulations were carried out in Matlab\cite{MATLAB:2024b}  using a simple 2-spin model to numerically support and describe Bayesian optimization for the trityl sample. The simulations involved density matrix evaluations under the influence of the Hamiltonian in Eq. (\ref{eq:2spins}). The offset $\Delta \omega_S$ was a variable parameter within the range of offsets chosen in the experiments. The proton Larmor frequency, $\omega_{0I}/(2\pi)$ corresponding to operation at X-band $\approx$ 14.8 MHz, was set by the experiment calibration. The secular and pseudo-secular hyperfine coupling constants $A$ and $B$, respectively, were calculated in a point-dipole model with an electron-proton distance of $r$=4.5 Å and a polar angle of $\beta$ = 45$\degree$. No powder averaging was used, as only a proportional polarization transfer is needed, as elaborated on in \cite{DNPbridge,SIMPSON_PP}. The formulas used for the hyperfine coupling constants are
    $A$  =-$\frac{\mu_0}{4\pi} \frac{\hbar\gamma_S\gamma_I}{r^3} (3 \cos^2(\beta)-1)$ and 
    $B$  =-$\frac{\mu_0}{4\pi} \frac{\hbar\gamma_S\gamma_I}{r^3} \frac{3}{2}\sin(2\beta)$,
where $\mu_0$ is the vacuum permeability, $\hbar$ is the reduced Planck constant, $\gamma_S$ and $\gamma_I$ are gyromagnetic ratios for electron and proton, respectively, $r$ is the electron-proton distance, and $\beta$ is the polar angle between them. The MW irradiation $\mathcal{H}_{MW}(t)$ was provided as a piecewise function using the amplitude, phase, and duration parameters for each experimental sequence. To translate the experimental MW amplitudes into frequencies, the maximal nutation frequency was measured and used as a reference for linear scaling. This maximal nutation was assumed constant for each optimization, but as seen previously from the nutation spectra, this is not always the case.\cite{GUPTA201517} Accordingly, the simulations in Figs. \ref{fig:FlipOpt}e and \ref{fig:FlipOpt}f used microwave (MW) inhomogeneity averaging using the profile in Supplementary Information Table 
S17.

\section*{Acknowledgments}
We acknowledge advice from Dr.\ A.\ Doll (Paul Scherrer Institut, CH) and Dr.\ D.\ Klose  and Prof.\ G.\ Jeschke (ETH, Z\"urich, CH) on building the pulsed X-band EPR/DNP instrumentation.
 The authors acknowledge  support from the Villum Foundation Synergy programme (grant 50099),  the Novo Nordisk Foundation (NERD grant NNF22OC0076002),  and the DeiC National HPC (g.a. DeiC-AU-N5-2025153)
 

 \paragraph*{Competing interests:}
There are no competing interests to declare.

\subsection*{Supplementary materials}
Supplementary Text\\
Figs. S1 to S16\\
Tables S1 to S17\\
\bibliography{science_template} 

@article{Bloch,
	author = {Bloch, F.},
	journal = {Phys. Rev.},
	month = {Oct},
	pages = {460--474},
	title = {Nuclear Induction},
	volume = {70},
	year = {1946}}

@article{Hahn,
	author = {Hahn, E. L.},
	journal = {Phys. Rev.},
	month = {Nov},
	pages = {580--594},
	title = {Spin Echoes},
	volume = {80},
	year = {1950}}

@article{Ramsey_1950,
	author = {Ramsey, Norman F.},
	journal = {Phys. Rev.},
	month = {Jun},
	pages = {695--699},
	title = {A Molecular Beam Resonance Method with Separated Oscillating Fields},
	volume = {78},
	year = {1950}}

@article{Hartmann_photon_echo_1964,
	author = {Kurnit, N. A. and Abella, I. D. and Hartmann, S. R.},
	journal = {Phys. Rev. Lett.},
	month = {Nov},
	pages = {567--568},
	title = {Observation of a Photon Echo},
	volume = {13},
	year = {1964}}

@article{Freed_coherent_EPR,
	author = {Freed, Jack H.},
	journal = {The Journal of Chemical Physics},
	month = {10},
	number = {7},
	pages = {2312-2332},
	title = {Theory of Saturation and Double‐Resonance Effects in ESR Spectra},
	volume = {43},
	year = {1965}}

@book{abragam1961principles,
	author = {Abragam, A.},
	publisher = {Clarendon Press},
	series = {International series of monographs on physics},
	title = {The Principles of Nuclear Magnetism},
	year = {1961}}

@book{Ernst-book,
	author = {Ernst, Richard R and Bodenhausen, Geoffrey and Wokaun, Alexander},
	month = {05},
	publisher = {Oxford University Press},
	title = {Principles of Nuclear Magnetic Resonance in One and Two Dimensions},
	year = {1990}}

@book{Jeschke_book,
	author = {Schweiger, Arthur and Jeschke, Gunnar},
	publisher = {Oxford university press},
	title = {Principles of pulse electron paramagnetic resonance},
	year = {2001}}

@article{Doll:2017wb,
	author = {Doll, Andrin and Jeschke, Gunnar},
	journal = {J Magn Reson},
	month = {Jul},
	pages = {46--62},
	title = {Wideband frequency-swept excitation in pulsed EPR spectroscopy.},
	volume = {280},
	year = {2017}}

@article{PLATO_adv,
	author = {Nielsen, A. B. and Carvalho, J. P. A. and Goodwin, D. L. and Wili, N. and Nielsen, N. C.},
	journal = {Phys. Chem. Chem. Phys.},
	pages = {28208-28219},
	title = {Dynamic nuclear polarization pulse sequence engineering using single-spin vector effective Hamiltonians},
	volume = {26},
	year = {2024}}

@article{rCW_EEHT_DNP,
	author = {Nielsen, Anders B. and Carvalho, Jos{\'e} P. and Wili, Nino and Jensen, Filip V. and Goodwin, David L. and Untidt, Thomas S. and To{\v s}ner, Zden{\v e}k and Nielsen, Niels Chr.},
	journal = {The Journal of Chemical Physics},
	month = {10},
	number = {14},
	pages = {144111},
	title = {Controlling effective Hamiltonians: Broadband pulsed dynamic nuclear polarization by constrained random walk and non-linear optimization},
	volume = {163},
	year = {2025}}

@article{shankar,
	author = {Shankar,Ravi and Ernst,Matthias and Madhu,P. K. and Vosegaard,Thomas and Nielsen,Niels Chr. and Nielsen,Anders B.},
	journal = {The Journal of Chemical Physics},
	number = {13},
	pages = {134105},
	title = {A general theoretical description of the influence of isotropic chemical shift in dipolar recoupling experiments for solid-state NMR},
	volume = {146},
	year = {2017}}

@article{SSV-EHT,
	author = {Nielsen,Anders B. and Hansen,Michael Ryan and Andersen,J{\o}rgen Ellegaard and Vosegaard,Thomas},
	journal = {The Journal of Chemical Physics},
	number = {13},
	pages = {134117},
	title = {Single-spin vector analysis of strongly coupled nuclei in TOCSY NMR experiments},
	volume = {151},
	year = {2019}}

@article{SVEHT_EEHT,
	author = {Anders B. Nielsen and Niels Chr. Nielsen},
	journal = {Journal of Magnetic Resonance Open},
	pages = {100064},
	title = {Accurate analysis and perspectives for systematic design of magnetic resonance experiments using single-spin vector and exact effective Hamiltonian theory},
	volume = {12-13},
	year = {2022}}

@article{AHT,
	author = {Haeberlen, U. and Waugh, J. S.},
	journal = {Phys. Rev.},
	month = {Nov},
	pages = {453--467},
	title = {Coherent Averaging Effects in Magnetic Resonance},
	volume = {175},
	year = {1968}}

@article{scBCH,
	author = {Hohwy,M. and Nielsen,N. C.},
	journal = {The Journal of Chemical Physics},
	number = {10},
	pages = {3780-3791},
	title = {Systematic design and evaluation of multiple-pulse experiments in nuclear magnetic resonance spectroscopy using a semi-continuous Baker--Campbell--Hausdorff expansion},
	volume = {109},
	year = {1998}}

@article{EEHT,
	author = {Untidt, Thomas S. and Nielsen, Niels Chr.},
	journal = {Phys. Rev. E},
	month = {Jan},
	pages = {021108},
	title = {Closed solution to the Baker-Campbell-Hausdorff problem: Exact effective Hamiltonian theory for analysis of nuclear-magnetic-resonance experiments},
	volume = {65},
	year = {2002}}

@article{EEHT2,
	author = {Siminovitch,David and Untidt,Thomas and Nielsen,Niels Chr.},
	journal = {The Journal of Chemical Physics},
	number = {1},
	pages = {51-66},
	title = {Exact effective Hamiltonian theory. II. Polynomial expansion of matrix functions and entangled unitary exponential operators},
	volume = {120},
	year = {2004}}

@article{MaximovSmoothing,
	author = {Maximov,Ivan I. and Salomon,Julien and Turinici,Gabriel and Nielsen,Niels Chr.},
	journal = {The Journal of Chemical Physics},
	number = {8},
	pages = {084107},
	title = {A smoothing monotonic convergent optimal control algorithm for nuclear magnetic resonance pulse sequence design},
	volume = {132},
	year = {2010}}

@article{OC_DNP,
	author = {Carvalho, Jos{\'e} P. and Goodwin, David L. and Wili, Nino and Nielsen, Anders Bodholt and Nielsen, Niels Chr.},
	journal = {The Journal of Chemical Physics},
	month = {02},
	number = {5},
	pages = {054111},
	title = {Optimal control design strategies for pulsed dynamic nuclear polarization},
	volume = {162},
	year = {2025}}

@article{GOODWIN_feedback_epr,
	author = {David L. Goodwin and William K. Myers and Christiane R. Timmel and Ilya Kuprov},
	journal = {Journal of Magnetic Resonance},
	pages = {9-16},
	title = {Feedback control optimisation of ESR experiments},
	volume = {297},
	year = {2018}}

@article{Chirp_ENDOR_mr-6-33-2025,
	author = {Stropp, J. and Wili, N. and Nielsen, N. C. and Klose, D.},
	journal = {Magnetic Resonance},
	number = {1},
	pages = {33--42},
	title = {Increased sensitivity in electron--nuclear double resonance spectroscopy with chirped radiofrequency pulses},
	volume = {6},
	year = {2025}}

@article{DNP_echo,
	author = {Nino Wili and Anders B. Nielsen and Jos{\'e} P. Carvalho and Niels Chr. Nielsen},
	journal = {Science Advances},
	number = {42},
	pages = {eadr2420},
	title = {Observation of dynamic nuclear polarization echoes},
	volume = {10},
	year = {2024}}

@article{grape,
	author = {Navin Khaneja and Timo Reiss and Cindie Kehlet and Thomas Schulte-Herbr{\"u}ggen and Steffen J. Glaser},
	journal = {Journal of Magnetic Resonance},
	number = {2},
	pages = {296-305},
	title = {Optimal control of coupled spin dynamics: design of NMR pulse sequences by gradient ascent algorithms},
	volume = {172},
	year = {2005}}

@article{MaximovOC,
	author = {Maximov,Ivan I. and To{\v s}ner,Zden{\u e}k and Nielsen,Niels Chr.},
	journal = {The Journal of Chemical Physics},
	number = {18},
	pages = {184505},
	title = {Optimal control design of NMR and dynamic nuclear polarization experiments using monotonically convergent algorithms},
	volume = {128},
	year = {2008}}

@article{SIMPSONOC,
	author = {Zden{\v e}k To{\v s}ner and Thomas Vosegaard and Cindie Kehlet and Navin Khaneja and Steffen J. Glaser and Niels Chr. Nielsen},
	journal = {Journal of Magnetic Resonance},
	number = {2},
	pages = {120-134},
	title = {Optimal control in NMR spectroscopy: Numerical implementation in SIMPSON},
	volume = {197},
	year = {2009}}

@article{OCKuprov,
	author = {Goodwin,D. L. and Kuprov,Ilya},
	journal = {The Journal of Chemical Physics},
	number = {20},
	pages = {204107},
	title = {Modified Newton-Raphson GRAPE methods for optimal control of spin systems},
	volume = {144},
	year = {2016}}

@article{Marshall_DNP_insitu,
	author = {Marshall, Alastair and Reisser, Thomas and Rembold, Phila and M\"uller, Christoph and Scheuer, Jochen and Gierse, Martin and Eichhorn, Tim and Steiner, Jakob M. and Hautle, Patrick and Calarco, Tommaso and Jelezko, Fedor and Plenio, Martin B. and Montangero, Simone and Schwartz, Ilai and M\"uller, Matthias M. and Neumann, Philipp},
	journal = {Phys. Rev. Res.},
	month = {Dec},
	pages = {043179},
	title = {Macroscopic hyperpolarization enhanced with quantum optimal control},
	volume = {4},
	year = {2022}}

@article{Mathies:2016aa,
	author = {Mathies, Guinevere and Jain, Sheetal and Reese, Marcel and Griffin, Robert G.},
	journal = {The Journal of Physical Chemistry Letters},
	month = {01},
	number = {1},
	pages = {111--116},
	title = {Pulsed Dynamic Nuclear Polarization with Trityl Radicals},
	volume = {7},
	year = {2016}}

@article{Pulsepol,
	author = {Ilai Schwartz and Jochen Scheuer and Benedikt Tratzmiller and Samuel M{\"u}ller and Qiong Chen and Ish Dhand and Zhen-Yu Wang and Christoph M{\"u}ller and Boris Naydenov and Fedor Jelezko and Martin B. Plenio},
	journal = {Science Advances},
	number = {8},
	pages = {eaat8978},
	title = {Robust optical polarization of nuclear spin baths using Hamiltonian engineering of nitrogen-vacancy center quantum dynamics},
	volume = {4},
	year = {2018}}

@article{TOP_DNP,
	author = {Kong Ooi Tan and Chen Yang and Ralph T. Weber and Guinevere Mathies and Robert G. Griffin},
	journal = {Science Advances},
	number = {1},
	pages = {eaav6909},
	title = {Time-optimized pulsed dynamic nuclear polarization},
	volume = {5},
	year = {2019}}

@article{XiX_DNP,
	author = {Redrouthu, Venkata SubbaRao and Mathies, Guinevere},
	journal = {Journal of the American Chemical Society},
	number = {4},
	pages = {1513-1516},
	title = {Efficient Pulsed Dynamic Nuclear Polarization with the X-Inverse-X Sequence},
	volume = {144},
	year = {2022}}

@article{TPPM_DNP,
	author = {Redrouthu, Venkata SubbaRao and Vinod-Kumar, Sanjay and Mathies, Guinevere},
	journal = {The Journal of Chemical Physics},
	month = {07},
	number = {1},
	pages = {014201},
	title = {{Dynamic nuclear polarization by two-pulse phase modulation}},
	volume = {159},
	year = {2023}}

@article{DNP_steady_state,
	author = {Jegadeesan, Shebha Anandhi and Zhao, Yujie and Smith, Graham M. and Kuprov, Ilya and Mathies, Guinevere},
	journal = {The Journal of Chemical Physics},
	month = {07},
	number = {3},
	pages = {034111},
	title = {Simulation of pulsed dynamic nuclear polarization in the steady state},
	volume = {163},
	year = {2025}}

@article{carvalho2026,
	author    = {Jos{\'e} P. Carvalho and Anders Bodholt Nielsen and David L. Goodwin and Nino Wili and Niels Chr. Nielsen},
	title     = {Longitudinal Pulsed Dynamic Nuclear Polarization Transfer via Periodic Optimal Control},
    	journal = {J. Phys. Chem. Lett.},
	note    = {In press},
	year      = {2026},
}

@article{BEAM,
	author = {Nino Wili and Anders Bodholt Nielsen and Laura Alicia V{\"o}lker and Lukas Schreder and Niels Chr. Nielsen and Gunnar Jeschke and Kong Ooi Tan},
	journal = {Science Advances},
	number = {28},
	pages = {eabq0536},
	title = {Designing broadband pulsed dynamic nuclear polarization sequences in static solids},
	volume = {8},
	year = {2022}}

@article{plato,
  title = {Dynamic Nuclear Polarization Pulse Sequence Engineering Using Single-Spin Vector Effective {{Hamiltonians}}},
  shorttitle = {{{PLATO}}},
  author = {Nielsen, A. B. and Carvalho, J. P. A. and Goodwin, D. L. and Wili, N. and Nielsen, N. C.},
  date = {2024},
  journaltitle = {Physical Chemistry Chemical Physics},
  shortjournal = {Phys. Chem. Chem. Phys.},
  volume = {26},
  number = {44},
  pages = {28208--28219},
  langid = {english}
}

@misc{Bayes,
      title={A Tutorial on Bayesian Optimization}, 
      author={Peter I. Frazier},
      year={2018},
      eprint={1807.02811},
      archivePrefix={arXiv},
      primaryClass={stat.ML},
      url={https://arxiv.org/abs/1807.02811}, 
}

@inproceedings{BayesKernel,
	author = {Snoek, Jasper and Larochelle, Hugo and Adams, Ryan P},
	booktitle = {Advances in Neural Information Processing Systems},
	editor = {F. Pereira and C.J. Burges and L. Bottou and K.Q. Weinberger},
	publisher = {Curran Associates, Inc.},
	title = {Practical Bayesian Optimization of Machine Learning Algorithms},
	volume = {25},
	year = {2012}}

@article{novel,
	author = {A Henstra and P Dirksen and J Schmidt and W.Th Wenckebach},
	journal = {Journal of Magnetic Resonance (1969)},
	number = {2},
	pages = {389-393},
	title = {Nuclear spin orientation via electron spin locking (NOVEL)},
	volume = {77},
	year = {1988}}

@article{SE,
  title = {Polarization of {{Nuclei}} by {{Resonance Saturation}} in {{Paramagnetic Crystals}}},
  shorttitle = {Solid-{{Effect}}},
  author = {Jeffries, C. D.},
  date = {1957},
  journaltitle = {Physical Review},
  shortjournal = {Phys. Rev.},
  volume = {106},
  number = {1},
  pages = {164--165},
  langid = {english}
}

@article{offNovel,
  title={Off-resonance NOVEL},
  author={Jain, Sheetal K and Mathies, Guinevere and Griffin, Robert G},
  journal={The Journal of chemical physics},
  volume={147},
  number={16},
  year={2017},
  publisher={AIP Publishing}
}

@ARTICLE{BayesDesign,
  author={Greenhill, Stewart and Rana, Santu and Gupta, Sunil and Vellanki, Pratibha and Venkatesh, Svetha},
  journal={IEEE Access}, 
  title={Bayesian Optimization for Adaptive Experimental Design: A Review}, 
  year={2020},
  volume={8},
  number={},
  pages={13937-13948}
}

@article{DNPbridge,
author = {Carvalho, Jos{\'e} P. and Nielsen, Anders Bodholt and Baligács, Enikő and Wili, Nino and Nielsen, Niels Chr.},
title = {Bridging Dynamic Nuclear Polarization and Solid-State NMR Dipolar Recoupling: From Static Single Crystal to Spinning Powders},
journal = {The Journal of Physical Chemistry Letters},
volume = {16},
number = {17},
pages = {4363-4371},
year = {2025}
}

@book{rasmussen,
	author = {Rasmussen, Carl Edward and Williams, Christopher K. I.},
	month = {11},
	publisher = {The MIT Press},
	title = {Gaussian Processes for Machine Learning},
	year = {2005}}

@misc{MatlabAcFunc,
  title = {Matlab documentation on Bayesian Optimization Algorithm},
  howpublished = {\url{https://se.mathworks.com/help/stats/bayesian-optimization-algorithm.html#bvaz8tr-1}}
}

@misc{AcFunc1,
      title={Convergence rates of efficient global optimization algorithms}, 
      author={Adam D. Bull},
      year={2011},
}

@misc{AcFunc2,
      title={Bayesian Optimization with Unknown Constraints}, 
      author={Michael A. Gelbart and Jasper Snoek and Ryan P. Adams},
      year={2014},
}

@misc{SIMPSON_PP,
	archiveprefix = {arXiv},
	author = {David L. Goodwin and Jose P. Carvalho and Anders B. Nielsen and Nino Wili and Andreas Brinkmann and Thomas Vosegaard and Zdenek Tosner and Niels Chr. Nielsen},
	eprint = {2602.15793},
	primaryclass = {physics.chem-ph},
	title = {Extending numerical simulations in SIMPSON: Electron paramagnetic resonance, dynamic nuclear polarisation, propagator splitting, pulse transients, and quadrupolar cross terms},
	url = {https://arxiv.org/abs/2602.15793},
	year = {2026}}

@book{press2007numerical,
	author = {Press, W.H. and Teukolsky, S.A. and Vetterling, W.T. and Flannery, B.P.},
	edition = 3,
	publisher = {Cambridge University Press},
	title = {Numerical Recipes: The Art of Scientific Computing},
	year = 2007}

@article{simplex,
	author = {Nelder, J. A. and Mead, R.},
	journal = {The Computer Journal},
	month = {01},
	number = {4},
	pages = {308-313},
	title = {{A Simplex Method for Function Minimization}},
	volume = {7},
	year = {1965}}

@article{GUPTA201517,
	author = {Rupal Gupta and Guangjin Hou and Tatyana Polenova and Alexander J. Vega},
	journal = {Solid State Nuclear Magnetic Resonance},
	pages = {17-26},
	title = {RF inhomogeneity and how it controls CPMAS},
	volume = {72},
	year = {2015}}

@book{MATLAB:2024b,
	address = {Natick, Massachusetts},
	author = {MATLAB},
	publisher = {The MathWorks Inc.},
	title = {version 24.2.0 (R2024b)},
	year = {2024}}
\bibliographystyle{sciencemag}




\newpage


\clearpage
\newpage


\renewcommand{\thefigure}{S\arabic{figure}}
\renewcommand{\thetable}{S\arabic{table}}
\renewcommand{\theequation}{S\arabic{equation}}
\renewcommand{\thepage}{S\arabic{page}}
\setcounter{figure}{0}
\setcounter{table}{0}
\setcounter{equation}{0}
\setcounter{page}{1} 
\renewcommand{\thesection}{S\arabic{section}}
\renewcommand{\thesubsection}{\arabic{subsection}}
\renewcommand{\thesubsubsection}{\arabic{subsection}.\alph{subsubsection}}
\renewcommand{\theparagraph}{\arabic{subsection}.\alph{subsubsection}.\arabic{paragraph}}

\onecolumn
\begin{center}
\section*{Supplementary Information for\\ \scititle}

Filip V. Jensen, 
José P. Carvalho, 
Nino Wili, 
Asbj{\o}rn Holk Thomsen, 
David L. Goodwin, 
Lukas Trottner, \\
Claudia Strauch, 
Anders Bodholt Nielsen$^{\ast}$, 
Niels Chr. Nielsen$^{\ast}$\\
\small$^\ast$Corresponding authors. Email: ncn@chem.au.dk abn@chem.au.dk
\end{center}

\subsubsection*{This PDF file includes:}
Supplementary Text\\
Supplementary Figs. S1 to S16\\
Supplementary Tables S1 to S17\\


\newpage
\setcounter{section}{0}
\renewcommand{\thesection}{S\arabic{section}}

\section{Supplementary Text}
\label{S1}

The following supplementary text provides more detailed information on the implementation of the Bayesian optimization integrated with the home-built pulsed DNP spectrometer (see main text) via the feedback-loop control scheme in Fig. 5 (main text) and the constrained random walk (cRW)\cite{rCW_EEHT_DNP} used to constrain the Bayesian operational space. Furthermore, the supplementary material (text, figures, and tables) provides details on calibrations, optimizations, parameters, and supplementary numerical Bayesian optimizations (\textit{in silico}, mimicking \textit{in situ} experimental procedures). The MATLAB \cite{MATLAB:2024b} code used for \textit{in silico} Bayesian optimization is also provided.

\subsection{Bayesian implementation for DNP optimizations}
\label{S11}

This section describes the Bayesian procedures, including variable definitions and the options for cRW constraints. The options are integral in the provided Bayesian optimization code and descriptive for the performed experimental and numerical optimizations as described in Supplementary Figures and Supplementary Tables.

\subsubsection{Bayesian algorithm in Matlab}
\label{S11a}
To perform Bayesian optimization for efficient DNP transfer, we have in this study used the Bayesian optimization package \matlab{bayesopt} implemented in the Matlab programming environment.\cite{MATLAB:2024b}  The Bayesian was implemented with the following instruction:

\begin{lstlisting}[style=Matlab-editor]
bayesopt(objectiveFunction ...
    ,variableDescriptions ...
    ,'AcquisitionFunctionName',options.acFunc ...
    ,'IsObjectiveDeterministic',options.deterministic ...
    ,'NumSeedPoints',4 ...
    ,'MaxObjectiveEvaluations',MaxEvals);
\end{lstlisting}
where the arguments (in red) and general settings for the Bayesian optimization are defined below.

\paragraph{Defining \matlab{objectiveFunction} in MATLAB for DNP optimization} 
\label{S11a1}~\\
The objective function for the Bayesian optimization is developed to enable communication with the spectrometer via the feedback-loop control procedure in Fig. 5 (main text). It receives control variables $x$ from \matlab{bayesopt}, runs the experiments, saves the experiment settings and results to the experiment log, and calculates the absolute mean of enhancements.

\begin{lstlisting}[style=Matlab-editor]
objectiveFunction = @(x) -abs(mean(bayesopt_wrapper(x,options)));
\end{lstlisting}
inside \matlab{bayesopt_wrapper} the control variables are used to determine the amplitude and phase modulations of the MW irradiation based on \matlab{options}. It then runs the pulse sequence for the chosen offsets and outputs DNP enhancements as a vector
\\
\begin{lstlisting}[style=Matlab-editor]
function [enhancements] = bayesopt_wrapper(x,options)
% Determine DNP sequence from variables
[excitationSeq,transferSeq] = DNP_sequence(x,options);
% Run DNP experiments and save to file
filepath = run_DNP(excitationSeq,transferSeq,options);
% Evaluate enhancements
reference = process_reference();
enhancements = get_DNP_enhancement(filepath,reference);
end
\end{lstlisting}

\paragraph{Definition of \matlab{options} in MATLAB for DNP optimization} 
\label{sec:options}~\\
The implementation of \matlab{bayesopt} is controlled via a set of \matlab{options}. Here a short explanation for each option is given, and in \Cref{tab:insito2ac,tab:Fig2e,tab:firstoptions,tab:insilicoS7,tab:insitoS13,tab:AmpAcFirst,tab:insilicoS5,tab:insilicoS75b,tab:AmpAcLast,tab:Fig3a,tab:insito4S15,tab:lastoptions,tab:insito4S15b},
the settings of these options are listed for each of the presented optimizations.

\matlab{options.acFunc} specifies the acquisition function $\mathrm{Ac}$ used in the Bayesian optimization. \matlab{bayesopt} has 4 built-in acquisition functions, discounting an acquisition function based on evaluation time: "expected-improvement", "expected-improvement-plus", "lower-confidence-bound" and "probability-of-improvement" as described in more detail in the underlying Matlab documentation.\cite{MATLAB:2024b,MatlabAcFunc,AcFunc1,AcFunc2} If acFunc is specified as “monte-carlo”, it implies that a Monte Carlo optimization was used instead of the Bayesian.

\matlab{options.deterministic} specifies if the objective function is deterministic. If set to "false", noise $\sigma$ is included in the Gaussian process model. This option was set to "true" for numerical optimizations and "false" for experimental optimizations.  Exceptions include the experimental optimizations in Figs. 2a-c and \ref{fig:PulseNumExp}, in which the objective function was treated as deterministic. One may, in retrospect, argue if this setting was optimal, but the low noise of the experimental model implies that this initial choice should not make a notable difference.

\matlab{options.nu1max} is the maximal MW amplitude $\omega_1/(2\pi)$ in MHz. For the experimental models, this value was determined from the resonator profile as shown in Fig. \ref{fig:ResProf}.

\matlab{options.nuI} is the nuclear Larmor $\omega_I/(2\pi)$ frequency in MHz.

\matlab{options.eOffsets} are the evaluated electron-spin offsets $\Delta \omega_S/(2\pi)$ in MHz. \matlab{bayesopt_wrapper} will output an enhancement for each of these offsets.

\matlab{options.transPulseNum} is the number of pulses in the transfer element.

\matlab{options.transAmpScale} is the range of possible MW amplitudes for pulses in a transfer element in units of maximal nutation frequency $\omega_{max}/(2\pi)$ before potential constraints are taken into account.

\matlab{options.transPhase} is the phase $\phi$ of the pulses in a transfer element in radians. If a range is specified, \matlab{bayesopt} optimizes the phase of each pulse individually within the given range.

\matlab{options.transPulseTime} is the duration of each pulse in a transfer element in ns.

\matlab{options.RWorder} is the order in which the accumulated nutation angles $\theta$ of the transfer element are fixed. "forward" fixes the angles in chronological order, "backtrack" fixes the angles in reverse order, and "factorize" fixes the angle that is most distant in time to currently fixed angles. These aspects are discussed in Supplementary Text Section \ref{S11b}
and visualized in Fig. \ref{fig:AngleOrder}.

\matlab{options.constrained} specifies if the transfer elements are constrained by a resonance condition. If set to "false", the Bayesian optimization will freely vary pulse amplitudes, phases and durations within the allowed ranges. If set to "true", the Bayesian will instead vary the parameters of a cRW and construct the transfer element as explained in Section \ref{S11b}. In the present work, only constraints for optimization of transfer elements on trityl OX063  were implemented. In principle other constraints could be derived and implemented in similar fashion.

\matlab{options.DNPres} is the DNP resonance chosen for the potential constraint, specified by "ZQ" or "DQ" and the value of $k_I$.

\matlab{options.DNPoffset} is the angular offset in degrees of the accumulated nutation angle with respect to the chosen DNP resonance. If a range is specified, the Bayesian optimization decides how close the transfer element should be to the resonance within the given range.

\vspace{1cm}
The following \matlab{options} are specific to the numerical simulations:

\matlab{options.excitation} defines amplitudes, phases and durations of the excitation sequence. It can be set to "ideal" to circumvent the initial excitation pulse and instead initialize the density matrix as $S_x$.

\matlab{options.maxContact} is the maximal contact time for the simulated transfer in ns. The simulation repeats the transfer element at zero electron-spin offset until a maximal proton polarization is found or until the maximal contact time is expended.

\matlab{options.epDist} is the distance between the electron and proton in Å.

\matlab{options.polAngles} is the number of polar angles used in the powder average, distributed equally from 0 to $\pi/2$ radians but omitting 0.

\matlab{options.MWinhom} specifies if the simulation should use MW amplitude inhomogeneity averaging given in Table \ref{tab:inhom}.

\vspace{1cm}
The following \matlab{options} are specific to the experimental implementation:

\matlab{options.B0} is the static magnetic field in T.

\matlab{options.nuS} is carrier frequency for the maximum of the electron resonance spectrum.

\matlab{options.NMRpulseTime} is the duration of the $\pi/2$ pulse in the solid-echo. In addition, the duration of the saturation pulses is set to 1.2 times this duration.

\matlab{options.excitePulseNum} is the number of pulses in the excitation sequence.

\matlab{options.exciteAmpScale} is the MW amplitude of the excitation pulses in units of maximal nutation frequency. If a range is specified, \matlab{bayesopt} optimizes the amplitude of each pulse individually within the given range.

\matlab{options.excitePhase} is the phase of the excitation pulses. If a range is specified, \matlab{bayesopt} optimizes the phase of each pulse individually within the given range.

\matlab{options.excitePulseTime} is the duration of each excitation pulse. If a range is specified, \matlab{bayesopt} optimizes the duration within the given range with all pulses having the same duration.

\matlab{options.transRepeat} is the number of repetitions of the transfer element within a single DNP contact. If a range is specified, \matlab{bayesopt} optimizes the number of repetitions within the given range.

\matlab{options.transContact} is the total contact time of the transfer sequence in ns and can be specified instead of \matlab{options.transPulseTime}. In this case the duration of the transfer pulses is set such that the contact time comes closest to the specified value with the given number of repetitions. If a range is specified, \matlab{bayesopt} optimizes the contact time within the given range.

\matlab{options.buildup} is the DNP buildup time $t_{\text{DNP}}$ for a single DNP experiment.

It is noted that in all cases where pulse durations were varied, the result was rounded to the nearest 0.125 ns, since this is the minimal digitization for the AWG (see Materials and Methods, main text) set to a 8 GHz sampling rate.

\paragraph{Defining \matlab{variableDescriptions} in MATLAB for feedback-loop \textit{in situ} DNP optimization}
\label{sec:variable}~\\
To run \matlab{bayesopt} a set of \matlab{variableDescriptions} have to be defined as a vector of the built-in function \matlab{optimizableVariable(variableName,range)}. It defines which values are allowed for each variable in $x$ and will therefore have the same number of elements. Its structure and ranges will, however, depend on the optimization setup.

If the contact time was set to be optimized, \matlab{optimizableVariable("t_contact",options.transContact)} was used as part of \matlab{variableDescriptions}. Likewise, if the repetitions of the transfer element were to be optimized, \matlab{optimizableVariable("n_rep",options.transRepeat,"Type","integer")} was used as part of \matlab{variableDescriptions}, where the \matlab{"Type","integer"} input specifies that the number of repetitions must be an integer. The rest of \matlab{variableDescriptions} describe the amplitudes and phases of the excitation and transfer pulses in the extent they were to be optimized.

For Bayesian optimizations without a resonance constraint, \matlab{variableDescriptions} is just defined as a vector containing the variables and their range defined by \matlab{options}. In this case the structure of the input vector $x$ will be
\[
x = \left[\omega_1^{(ex,1)}/\omega_{max},\phi^{(ex,1)},...,\omega_1^{(tr,1)}/\omega_{max},\phi^{(tr,1)},...,t_{contact},n_{rep}\right]
\]
where $\omega_1/(2\pi)$ is the MW nutation frequency, $\phi$  the phase, $t_{\text{contact}}$  the contact time, and $n_{\text{rep}}$  the number of transfer element repetitions. The superscripts $(ex,i)$ and $(tr,i)$ signify the $i$'th excitation and transfer pulse, respectively.

\subsubsection{Constrained random walk}
\label{S11b}

When the transfer element was constrained by a resonance condition, the \matlab{variableDescriptions} were instead used for a cRW. The cRW\cite{rCW_EEHT_DNP} describes the progression of accumulated nutation angles $\theta_0,...,\theta_p$ at time point $t_0,...,t_p$, where $p$ is the number of pulses in the transfer element, $t_0=0$ and $\Delta t = t_{i+1}-t_i$ is the duration of a single pulse, which was in our case chosen to be constant. When the cRW has been determined, the corresponding nutation frequencies of the pulses can be calculated using a difference quotient
\begin{equation}
    \omega_1^{(tr,i)} = \frac{\theta_i-\theta_{i-1}}{\Delta t} \quad \text{for} \quad i \in\{1,\dots,p\} \quad . \label{eq:PulseSeq}
\end{equation}

To construct the cRW, the accumulated nutation angles have to obey the constraints of maximal MW nutation frequency, $\omega_{max}$, and the DNP resonance,
\begin{align}
    \theta_0 & = 0\quad, \\
    \left|\frac{\theta_i-\theta_{j}}{t_i-t_{j}}\right| & \leq \omega_{max} \quad \text{for} \quad i \neq j \in\{0,\dots,p\} \quad, \label{eq:MWmax} \\ 
    \left|\theta_p - \theta^{\text{DNP}} \right| & \leq \Delta\quad, \label{eq:resRange}
\end{align}
where $\theta^{\text{DNP}}$ is the nutation angle at the DNP resonance and $\Delta$ is the maximal angular offset from resonance.

Initially, only $\theta_0$ is fixed while the rest of the angles can be set within a range
\begin{equation}
    \theta_i = (\theta_{i,max}-\theta_{i,min}) x_i+\theta_{i,min}\quad, \label{eq:RWminmax}
\end{equation}
where $x_i \in [0,1]$ is the $i$'th variable in $x$ and $\theta_{i,max}$ and $\theta_{i,min}$ are the maximal and minimal allowed angle for $\theta_i$, respectively.

\begin{figure}[ht]
    \centering \includegraphics[width=\linewidth]{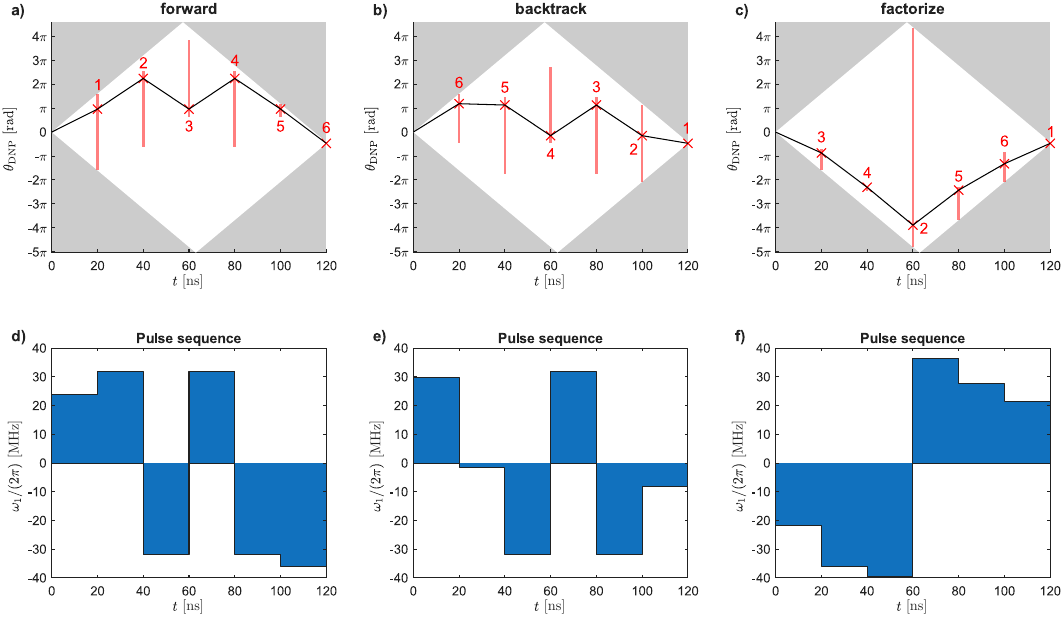} 
    \caption{(a-c) Three different ways of constructing the cRW. The red crosses show the resulting accumulated nutation angles. The numbers indicate the order in which the angles were fixed, and the vertical bars show the available range. All three cRW's used the input vector $x = [0.8,0.9,0.1,0.9,0.6,0.1]$, 6 pulses each of 20 ns duration, 14.8 MHz proton Larmor frequency, 40 MHz maximal nutation frequency, and the ZQ, $k_I=2$ DNP resonance with -5 to 5 degrees DNP offset. a) Angles are fixed in chronological order. b) Angles are fixed in reverse order. c) Angles that are most distant in time to any currently fixed angles are fixed first. (d-f) Corresponding pulse sequence for each of the cRW's given as $\pm x$-phase amplitudes.}
    \label{fig:AngleOrder}
\end{figure}

The allowed values for $\theta_i$ will depend on the value of nearby angles $\theta_j$ that have already been fixed as described by Eq. (\ref{eq:MWmax}). Therefore, the resulting cRW, and consequently the pulse sequence, will depend on the order in which the angles $\theta_i$ are set. This is illustrated in Fig. \ref{fig:AngleOrder} for the three different orders used in the present publication. "forward" fixes the angles in chronological order, "backtrack" fixes the angles in reverse order, and "factorize" fixes the angle that is most distant in time to the currently fixed angles. To calculate the allowed values for $\theta_i$, only the closest fixed angle to the left $\theta_{L(i)}$ and to the right $\theta_{R(i)}$ is needed
\begin{align}
    \theta_{i,max} & = \min\left\{ \theta_{L(i)}+\omega_{max} (t_i-t_{L(i)}),~ \theta_{R(i)}+\omega_{max} (t_{R(i)}-t_i) \right\} \label{eq:RW1} \\
    \theta_{i,min} & = \max\left\{ \theta_{L(i)}-\omega_{max} (t_i-t_{L(i)}),~ \theta_{R(i)}-\omega_{max} (t_{R(i)}-t_i) \right\} \label{eq:RW2} \quad .
\end{align}
If there are no fixed angles to the right of $\theta_i$, Eq. (\ref{eq:resRange}) must instead be taken into account
\begin{align}
    \theta_{i,max} & = \min\left\{ \theta_{L(i)}+\omega_{max} (t_i-t_{L(i)}),~ 
    \theta_{DNP}+\Delta+\omega_{max} (t_p-t_i) \right\} \label{eq:RW3} \\
    \theta_{i,min} & = \max\left\{ \theta_{L(i)}-\omega_{max} (t_i-t_{L(i)}),~ 
    \theta_{DNP}-\Delta-\omega_{max} (t_p-t_i) \right\} \quad .\label{eq:RW4}
\end{align}

\begin{figure}[ht]
    \centering \includegraphics[width=\linewidth]{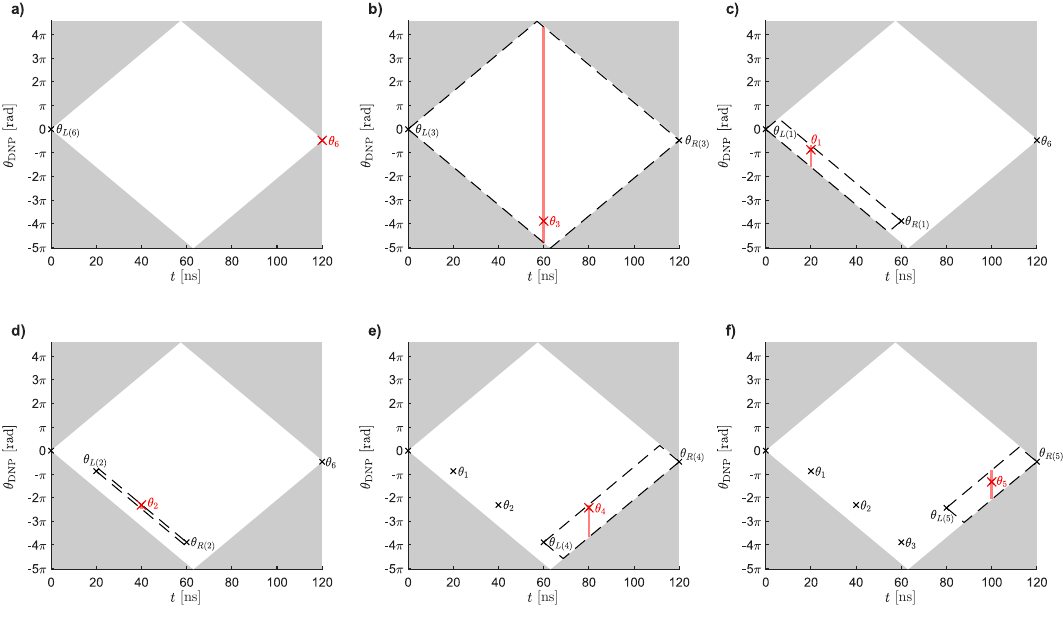} 
    \caption{A walkthrough of how to construct the cRW seen in Fig. \ref{fig:AngleOrder}c using the input vector $x = [0.8,0.9,0.1,0.9,0.6,0.1]$ and the "factorize" order where most distant angles are fixed first. In all panels the slope of the dashed line corresponds to maximal positive or negative nutation frequency of 40 MHz. a) Only the first angle is fixed and therefore $\theta_6$ is furthest away in time. Since it only has a fixed angle to the left $\theta_{L(6)}=\theta_0$, it uses Eqs. (\ref{eq:RW3}) - (\ref{eq:RW4}) to determine the allowed range (only $10\degree$ wide and too small to see on figure). The angle $\theta_6$ is placed 10\% up this range according to $x_6=0.1$. b) The first and last angles are fixed so $\theta_3$ is furthest away. The nearest fixed angles are $\theta_{L(3)}=\theta_0$ and $\theta_{R(3)}=\theta_6$ and Eqs. (\ref{eq:RW1}) - (\ref{eq:RW2}) determine the allowed range (red line). The angle $\theta_3$ is placed 10\% up this line according to $x_3=0.1$. c) All undetermined angles are 20 ns from a fixed angle so in this case the first of them $\theta_1$ is chosen. The nearest fixed angles are $\theta_{L(1)}=\theta_0$ and $\theta_{R(1)}=\theta_3$ and Eqs. (\ref{eq:RW1}) - (\ref{eq:RW2}) determine the allowed range (red line). The angle $\theta_1$ is placed 80\% up this line according to $x_1=0.8$. d) Next $\theta_2$ is chosen. The nearest fixed angles are $\theta_{L(2)}=\theta_1$ and $\theta_{R(2)}=\theta_3$ and Eqs. (\ref{eq:RW1}) - (\ref{eq:RW2}) determine the allowed range (red line, barely visible). The angle $\theta_2$ is placed 90\% up this line according to $x_2=0.9$. e) Next $\theta_4$ is chosen. The nearest fixed angles are $\theta_{L(4)}=\theta_3$ and $\theta_{R(4)}=\theta_6$ and Eq. (\ref{eq:RW1}) - (\ref{eq:RW2}) determine the allowed range (red line). The angle $\theta_4$ is placed 90\% up this line according to $x_4=0.9$. f) The last angle $\theta_5$ is chosen. The nearest fixed angles are $\theta_{L(5)}=\theta_4$ and $\theta_{R(5)}=\theta_6$ and Eqs. (\ref{eq:RW1}) - (\ref{eq:RW2}) determine the allowed range (red line). The angle $\theta_5$ is placed 60\% up this line according to $x_5=0.6$.}
    \label{fig:AngleOrder2}
\end{figure}

To illustrate how this works, Fig. \ref{fig:AngleOrder2} walks through the process of constructing the cRW shown in Fig. \ref{fig:AngleOrder}c. From Eqs. (\ref{eq:RWminmax}-\ref{eq:RW4}), the accumulated nutation angles can be determined from the input vector $x$ and the \matlab{options} which then gives the pulse sequence by Eq. (\ref{eq:PulseSeq}). In the end the structure of the input vector $x$ for optimization of a $p$-pulse constrained transfer element was
\[
x = \left[x_1,...,x_p,n_{\text{rep}}\right] \quad ,
\]
where the $n_{\text{rep}}$ variable is added only if the number of repetitions is being optimized.

The reason for defining these alternative cRW orders is that the straightforward approach of fixing the angles in chronological order often results in a final nutation angle near the boundaries of the DNP resonance in Eq. (\ref{eq:resRange}). The effect of choosing between the different RW orders and the different acquisition functions has been investigated numerically in Fig. \ref{fig:AmpAc}. From this figure, it was found that the "factorize" cRW order performed best and for the optimization in Fig. 2e (main text), this order was selected along with the acquisition function "expected-improvement-plus".

\subsubsection{Bayesian optimization MATLAB code}
\label{S11c}
In this section, we provide the Matlab code used for \textit{in silico} Bayesian optimization, noting that the corresponding experimental code depends specifically on the software steering our home-built pulsed DNP/EPR spectrometer and as such is not relevant to deploy generally. The presented code, however, illustrates all relevant aspects of the Bayesian optimization.
\begin{itemize}
\item Optimization setup script.
\begin{lstlisting}
%% Physical inputs
options.nuI = 14.8; % nuclear zeeman
options.nu1max = 32; % maximal irradiation strength
options.epDist=4.5; % proton-electron distance in Angstrom
%% Orientations and offsets
options.polarAngles=20; % Number of orientations from 0 to pi/2
options.eOffsets = -60:0.5:60; % electron offsets to be analyzed
options.MWinhom = false;
%% Flip definitions
options.excitation = 'ideal';
%% Transfer definitions
options.transPulseNum = 24;
options.transPulseTime = 5;
options.maxContact = 10e3;
options.DNPres = {'ZQ',2};
options.DNPoffset = 5;
options.RWorder = "forward";
%% bayesopt options
options.aqFunc = "expected-improvement-plus";
MaxEvals = 1000;
rng('shuffle')
%% Optimization
variableDescriptions = [];
for i = 1:options.transPulseNum
    variableDescriptions = [variableDescriptions,optimizableVariable...
    (['x' num2str(i)],[0,1])];
end
objectiveFunction = @(x) -abs(mean(numeric_bayesopt_wrapper(x,options)));
inputs = {objectiveFunction,variableDescriptions ...
    ,'AcquisitionFunctionName',options.aqFunc ...
    ,'IsObjectiveDeterministic',true ...
    ,'MaxObjectiveEvaluations',MaxEvals ...
    ,'Verbose',1 ...
    ,'PlotFcn',{} ...
    ,'OutputFcn',@saveToFile ...
    };
bayesopt(inputs{:});
\end{lstlisting}
\item Objective function for numerical implementation.
\begin{lstlisting}
function [output] = numeric_bayesopt_wrapper(x,options)
if istable(x)
    x = table2array(x);
end
if strcmp(options.excitation,'ideal')
    flip = zeros(3,0);
else
    flip = options.excitation;
end
transfer = construct_pulse_sequence(x,options);
output = numeric_DNP(flip,transfer,options);
end
\end{lstlisting}
\item Code for calculating transfer element.
\begin{lstlisting}
function [transfer] = transfer_element(x,options)
transfer = zeros(3,options.transPulseNum);
%% Determine pulsetime
transfer(3,:) = options.transPulseTime * ones(1,options.transPulseNum);
%% Find matching
effrot = options.nuI*sum(transfer(3,:))/1000;
if strcmp(options.DNPres{1},'ZQ')
    thetaDNP = 2*pi*(effrot-options.DNPres{2});
elseif strcmp(options.DNPres{1},'DQ')
    thetaDNP = -2*pi*(effrot-options.DNPres{2});
else
    error([options.DNPres{1} ' is not a valid transfer.'])
end
if abs(thetaDNP)-options.DNPoffset*pi/180>2*pi*options.nu1max*sum(transfer(3,:))
    error('requested overall flip angle physically impossible. lower kIa and/or change ZQ/DQ')
end
%% Generate cRW
theta = generate_cRW(x,thetaDNP+options.DNPoffset*pi/180*[-1,1],transfer(3,:),options.nu1max,options.RWorder,[-Inf,Inf]);
%% Calculate pulse sequence
amps = diff(theta)/(2*pi)./transfer(3,:)*1000/options.nu1max;
amps(isnan(amps)) = 0;
transfer(1,:) = abs(amps);
transfer(2,:) = pi*(1-sign(amps))/2;
end
\end{lstlisting}
\item Code for constructing cRW.
\begin{lstlisting}
    function [theta] = generate_cRW(x,target,pulsetimes,maxfreq,RWorder,MinMax)
%Convert from nanoseconds to microseconds
pulsetimes = pulsetimes/1000;
if isscalar(maxfreq)
    posfreq = maxfreq;
    negfreq = maxfreq;
else
    posfreq = maxfreq(2);
    negfreq = maxfreq(1);
end
%% Checks
if isscalar(target)
    target = [target,target];
end
if nargin < 5
    RWorder = "forward";
end
if nargin < 6
    MinMax = [-Inf,Inf];
end
if target(1) > target(2)
    error(['The target range [' num2str(target(1)) ',' num2str(target(2)) '] must be increasing.' ])
end
if MinMax(1) > MinMax(2)
    error(['The bounds [' num2str(MinMax(1)) ',' num2str(MinMax(2)) '] must be increasing.' ])
end
if MinMax(1) > 0 || MinMax(2) < 0
    error(['The starting point 0 must be within the bounds: [' num2str(MinMax(1)) ',' num2str(MinMax(2)) ']'])
end
if target(2) < MinMax(1) || target(1) > MinMax(2)
    error(['Cannot reach target range of [' num2str(target(1)) ','... 
    num2str(target(2)) '] within bounds: [' num2str(MinMax(1)) ',' num2str(MinMax(2)) ']'])
end
if target(2) < -2*pi*sum(pulsetimes)*negfreq || target(1) > 2*pi*sum(pulsetimes)*posfreq
    error(['Cannot reach target range of [' num2str(target(1)) ',' ...
    num2str(target(2)) '] rad within ' num2str(sum(pulsetimes)) '\mus with a maximal nutation of '...
 num2str(negfreq) '/' num2str(posfreq) ' MHz'])
end
%% Picking flip angles during the pulse sequence consistent with experiment options
pulsenum = length(pulsetimes);
theta = zeros(1,pulsenum+1);
maxi = MinMax(2)*ones(1,pulsenum);
mini = MinMax(1)*ones(1,pulsenum);
switch RWorder
    case 'forward'
        for j = 1:pulsenum
            maxi(j) = min([theta(j)+2*pi*posfreq*pulsetimes(j),target(2)+2*pi*negfreq*sum(pulsetimes(j+1:end)),maxi(j)]);
            mini(j) = max([theta(j)-2*pi*negfreq*pulsetimes(j),target(1)-2*pi*posfreq*sum(pulsetimes(j+1:end)),mini(j)]);
            theta(j+1) = mini(j) + x(j)*(maxi(j)-mini(j));
        end
    case 'backtrack'
        maxi(end) = min([2*pi*posfreq*sum(pulsetimes),target(2),maxi(end)]);
        mini(end) = max([-2*pi*negfreq*sum(pulsetimes),target(1),mini(end)]);
        theta(end) = mini(end) + x(end)*(maxi(end)-mini(end));
        for j = fliplr(1:pulsenum-1)
            maxi(j) = min([theta(j+2)+2*pi*posfreq*pulsetimes(j+1),2*pi*negfreq*sum(pulsetimes(1:j)),maxi(j)]);
            mini(j) = max([theta(j+2)-2*pi*negfreq*pulsetimes(j+1),-2*pi*posfreq*sum(pulsetimes(1:j)),mini(j)]);
            theta(j+1) = mini(j) + x(j)*(maxi(j)-mini(j));
        end
    case 'factorize'
        count = 1;
        maxi(end) = min([2*pi*posfreq*sum(pulsetimes),target(2),maxi(end)]);
        mini(end) = max([-2*pi*negfreq*sum(pulsetimes),target(1),mini(end)]);
        theta(end) = mini(end) + x(end)*(maxi(end)-mini(end));
        factors = factor(pulsenum);
        product = [1,cumprod(factors)];
        for j = 1:length(factors)
            secsize = pulsenum/product(j);
            partsize = secsize/factors(j);
            for sec = 1:product(j)
                endindex = secsize*sec+1;
                for part = 1:factors(j)-1
                    startindex = secsize*(sec-1)+(part-1)*partsize+1;
                    pulseindex = secsize*(sec-1)+part*partsize;
                    maxi(pulseindex) = min([theta(startindex) + 2*pi*posfreq*sum(pulsetimes(startindex:pulseindex)), ...
                        theta(endindex) + 2*pi*negfreq*sum(pulsetimes(pulseindex+1:endindex-1)), ...
                        maxi(pulseindex)]);
                    mini(pulseindex) = max([theta(startindex) - 2*pi*negfreq*sum(pulsetimes(startindex:pulseindex)), ...
                        theta(endindex) - 2*pi*posfreq*sum(pulsetimes(pulseindex+1:endindex-1)), ...
                        mini(pulseindex)]);
                    theta(pulseindex+1) = mini(pulseindex) + x(pulseindex)*(maxi(pulseindex)-mini(pulseindex));
                    count = count+1;
                end
            end
        end
    otherwise
        error(['Error in pulse sequence construction: ' RWorder ' is not a valid amplitude type.'])
end
end
\end{lstlisting}
\item Code for numerical simulation of DNP on a 2-spin system.
\begin{lstlisting}
function protonPolarization = numeric_DNP(flip,transfer,options)
%% Set polar angles and MW inhomogeneity
if options.polarAngles == 1
    pol_ang = pi/4;
else
    pol_ang = linspace(0,pi/2,options.polarAngles+1);
    pol_ang(1) = [];
end
if options.MWinhom
    inhom = [0.65 0.0484168
        0.70 0.052932
        0.75 0.0586464
        0.80 0.0662188
        0.85 0.0769463
        0.90 0.0938953
        0.95 0.12731
        1.00 0.32881
        1.05 0.146825];
else
    inhom = [1,1];
end
%% Physical operators and constants
X = 0.5*[0,1;
    1,0];
Y = 0.5i*[0,-1;
    1,0];
Z = 0.5*[1,0;
    0,-1];
Sx = kron(X,eye(2));
Sy = kron(Y,eye(2));
Sz = kron(Z,eye(2));
Ix = kron(eye(2),X);
Iz = kron(eye(2),Z);
if strcmp(options.excitation,'ideal')
    flip = zeros(3,0);
    rho0 = Sx; % Set starting state
else
    rho0 = Sz; % Set starting state
end
detect = Iz; % Set detection operator
mu0=1.256637062120000e-06;
gfree = 2.002319304362560;
bmagn = 9.274010078300000e-24;
g1H = 5.585694680000000;
planck = 6.626070150000000e-34;
nmagn = 5.050783746100000e-27;
T = mu0/(4*pi)*gfree*bmagn*g1H*nmagn*1/(options.epDist*1e-10)^3/planck/1e6;
%% Run flip and prepare transfer for zero offset
% scaling for different orientations
weights=sin(pol_ang); weights=weights/sum(weights);
buildup = 0;
rho = cell(options.polarAngles,size(inhom,1));
rho(:) = {rho0};
Uround = cell(options.polarAngles,size(inhom,1));
Uround(:) = {eye(size(Sz))};
for i = 1:options.polarAngles
    % Nuclear Larmor term and hyperfine coupling
    A=T*(3*cos(pol_ang(i))^2-1);
    B=3*sin(pol_ang(i))*cos(pol_ang(i))*T;
    H0 = options.nuI*Iz+A*Sz*Iz+B*Sz*Ix;
    for j = 1:size(inhom,1)
        for ip = 1:size(flip,2) % Run flip pulse sequence
            H1 = flip(1,ip)*inhom(j,1)*options.nu1max*( Sx*cos(flip(2,ip))+Sy*sin(flip(2,ip)) );
            % propagator for flip pulse
            Uflip = expm(-1i*2*pi*(H0+H1)*flip(3,ip)*1e-3);
            rho{i,j} = Uflip*rho{i,j}*Uflip';
        end
        buildup=buildup + weights(i)*inhom(j,2)*real(trace(detect*rho{i,j})/trace(detect*detect));
        % propagator for transfer element
        for ip = 1:size(transfer,2)
            H1 = transfer(1,ip)*inhom(j,1)*options.nu1max*( Sx*cos(transfer(2,ip))+Sy*sin(transfer(2,ip)) );
            Uround{i,j} = expm(-1i*2*pi*(H0+H1)*transfer(3,ip)*1e-3)*Uround{i,j};
        end
    end
end
count = 0;
while ~exist('nrep')
    count = count+1;
    buildup = [buildup,0];
    for i = 1:options.polarAngles
        for j = 1:size(inhom,1)
            rho{i,j} = Uround{i,j}*rho{i,j}*Uround{i,j}';
            buildup(end) = buildup(end) + weights(i)*inhom(j,2)*real(trace(detect*rho{i,j})/trace(detect*detect));
        end
    end
    if abs(buildup(end)) < buildup(end-1)
        nrep = count-1;
    elseif (count+1)*sum(transfer(3,:)) > options.maxContact
        [~,count] = max(abs(buildup));
        nrep = count-1;
    end
end
%% Run offset flip and transfer
profiles = zeros(length(options.eOffsets),nrep+1);
offzero = find(options.eOffsets == 0);
profiles(offzero,:) = buildup(1:nrep+1);
offind = 1:length(options.eOffsets);
offind(offzero) = [];
for i = 1:options.polarAngles
    % Nuclear Larmor and hyperfine coupling
    A=T*(3*cos(pol_ang(i))^2-1);
    B=3*sin(pol_ang(i))*cos(pol_ang(i))*T;
    H0 = options.nuI*Iz+A*Sz*Iz+B*Sz*Ix;
    for j = 1:size(inhom,1)
        for o = offind
            rho = rho0; % Set starting state
            for ip = 1:size(flip,2) % Run flip pulse sequence
                H1 = options.eOffsets(o)*Sz ...
                    + flip(1,ip)*inhom(j,1)*options.nu1max*( Sx*cos(flip(2,ip))+Sy*sin(flip(2,ip)) );
                % propagator for flip pulse
                Uflip = expm(-1i*2*pi*(H0+H1)*flip(3,ip)*1e-3);
                rho = Uflip*rho*Uflip';
            end
            % propagator for transfer element
            Uround = eye(size(Sz));
            for ip = 1:size(transfer,2)
                H1 = options.eOffsets(o)*Sz ...
                    + transfer(1,ip)*inhom(j,1)*options.nu1max*( Sx*cos(transfer(2,ip))+Sy*sin(transfer(2,ip)) );
                Uround = expm(-1i*2*pi*(H0+H1)*transfer(3,ip)*1e-3)*Uround;
            end
            for istep = 1:nrep
                profiles(o,istep)=profiles(o,istep) + weights(i)*inhom(j,2)*real(trace(detect*rho)/trace(detect*detect));
                rho = Uround*rho*Uround';
            end
            profiles(o,end)=profiles(o,end) + weights(i)*inhom(j,2)*real(trace(detect*rho)/trace(detect*detect));
        end
    end
end
protonPolarization = profiles(:,end)';
end
\end{lstlisting}
\end{itemize}
\subsection{Description of supplementary numerical optimizations}
\label{S12}

To investigate the performance of various implementations of Monte Carlo and Bayesian pulse sequence optimizations, we conducted a series of \textit{in silico} optimizations as illustrated in \Cref{fig:AmpAc,fig:PulseNum,fig:AcVar}. 

\subsubsection{\textit{In silico} pulse sequence optimizations}
\label{S12a}
More specifically, to compare with \textit{in situ} experimental optimizations and investigate the influence of different optimization procedures and number of pulses, we conducted the following \textit{in silico} optimizations. These were typically faster than the experimental optimizations and provided more direct information on specific determinants in the optimizations. 

\begin{itemize}

\item Fig. \ref{fig:AcVar}: Comparison of different optimization methods for DNP with 6-pulse elements to compare directly with the \textit{in situ} optimizations in Fig. 2 (main text). Qualitatively the \textit{in silico} and \textit{in situ} optimizations display similar trends, however, with the difference that the numerically optimized sequences are less broadbanded which may ascribed partly to that the \textit{in situ} optimizations automatically folds with the lineshape of the EPR line of  trityl OX063. We note that the cRW-OPT pulse sequences\cite{rCW_EEHT_DNP} numerically were better in this respect. The parameters used for the optimizations are given in \Cref{tab:firstoptions}.

\item Fig. \ref{fig:PulseNum}: Comparing cRW constrained Bayesian optimization with different numbers of pulses. As Bayesian optimization, in terms of speed and needed degrees of optimization freedom, depends on the number of variables in the optimization, we explored \textit{in silico} the performance of 1000 constrained Bayesian evaluations for cases of 6, 12, and 24 pulses. Inspecting, the transfer efficiency integrated over offset (as in Fig. 2, main text), it appears that 24 pulses challenge convergence while both 6 and 12 pulses provide reasonable efficiencies and offset profiles. We note that the experimental evaluation samples a larger spin system and more instrumental errors than included in the numerical (included MW inhomogeneity, see below) mimic. The parameters used for the optimizations are given in \Cref{tab:insilicoS7}.

\item Fig. \ref{fig:AmpAc}: Exploration of different acquisition functions and cRW settings. From this analysis, we established optimal parameters for the \textit{in situ}  optimizations as discussed in Supplementary Text Section \ref{sec:variable}. The parameters used for the optimizations are given in \Cref{tab:AmpAcFirst,tab:insilicoS5,tab:insilicoS75b,tab:AmpAcLast}.
\end{itemize}

\subsubsection{MW inhomogeneity profile}
\label{S12b}
To reproduce the best possible experimental conditions by numerical simulations, the simulations in Fig. 3e-f take into account an MW inhomogeneity profile as listed in Table. \ref{tab:inhom}. The specific profile was established in a power model as described by Gupta et al. \cite{GUPTA201517}

\subsection{Description of supplementary experimental data}
\label{S13}

To support the experimental data for Monte Carlo and Bayesian optimization, we present here supplementary data in form of calibrations, parameter evaluations, and additional optimizations. 

\subsubsection{Calibrations and background experiments}
\label{S13a}

To conduct \textit{in situ} optimization of broadband DNP experiments and compare with reference experiments, we initially calibrated parameters for the investigated samples on the DNP spectrometer. The parameters also served as input to numerical simulations.

\begin{itemize}
\item Fig. \ref{fig:ResProf}: Nutation curves (in time and frequency domain) and resonator profiles recorded as part of optimization of experimental parameters. 

\item Fig. \ref{fig:FieldSweeps}: EPR resonance profiles  recorded by field sweep for the two  trityl OX063 samples with difference solvent deuteration, as well as for 4-Oxo-TEMPO.

\item Fig. \ref{fig:T1}: Proton and electron T$_1$ relaxation profiles for the two trityl OX063 samples and 4-Oxo-TEMPO recorded as part of optimization of experimental parameters. 

\end{itemize}

\subsubsection{Pulse sequence optimizations and pulse sequence analysis}
\label{S13b}

\begin{figure}[ht]
    \centering
    \includegraphics[width=0.55\linewidth]{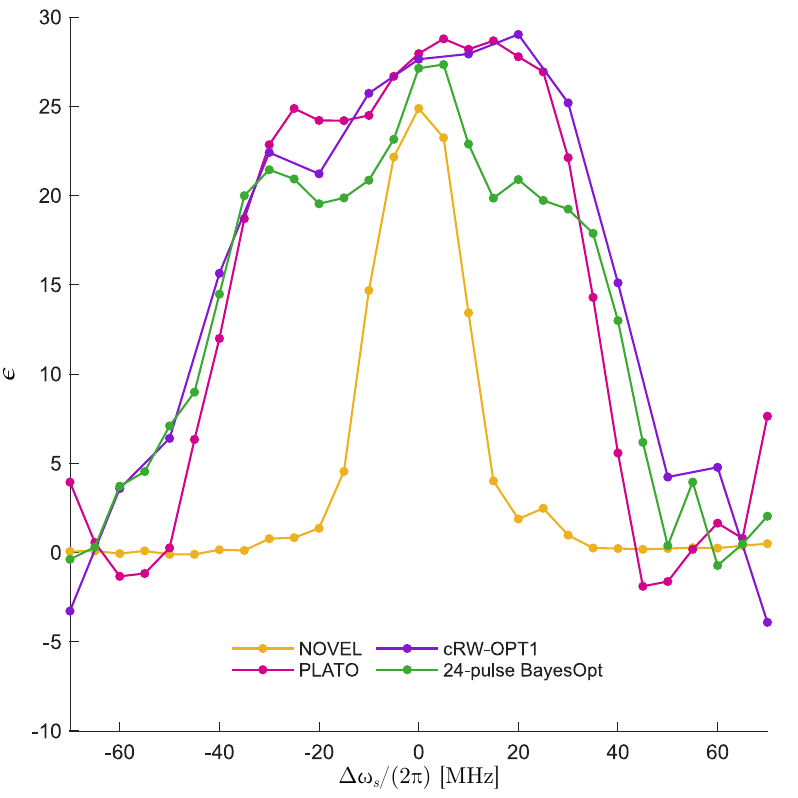} 
    \caption{Experimental \textit{in situ} polarization enhancements as a function of electron-spin offset using the same pulse sequences employed in Fig. 2f but on a sample of trityl in glycerol-d$_5$ and with 2 s buildup time.}
    \label{fig:OffsetProfiles-d5}
\end{figure}

\begin{figure}[ht]
    \centering
    \includegraphics[width=\linewidth]{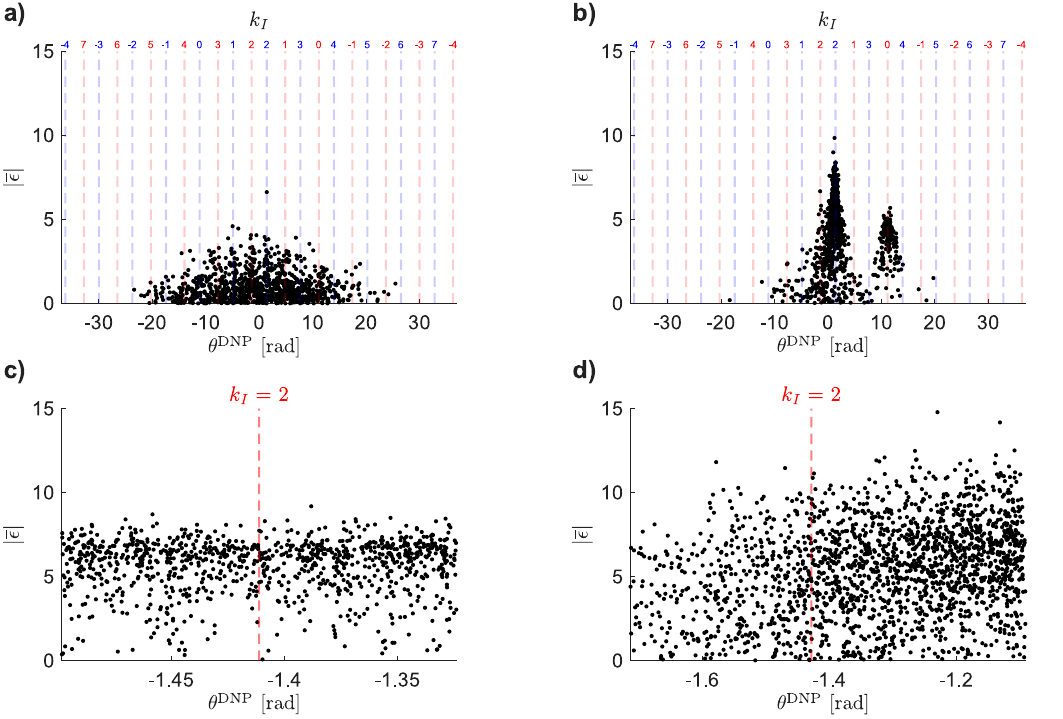} 
    \caption{Enhancement factors for the \textit{in situ} optimized DNP sequences in Figs. 2 a1-c1,e, as a function of the accumulated nutation angle $\theta^{DNP}$ of the transfer element for the given sequence. In each panel, the nutation angle axis spans the allowed range of effective angles which in the constrained optimizations centers around the selected resonance. The optimizations were: a) Monte Carlo, b) Bayesian, c) constrained Bayesian, and d) constrained Bayesian with refined options (see Table \ref{tab:Fig2e}).}
    \label{fig:Fig2EffField}
\end{figure}

\begin{figure}[ht]
    \centering
    \includegraphics[width=\linewidth]{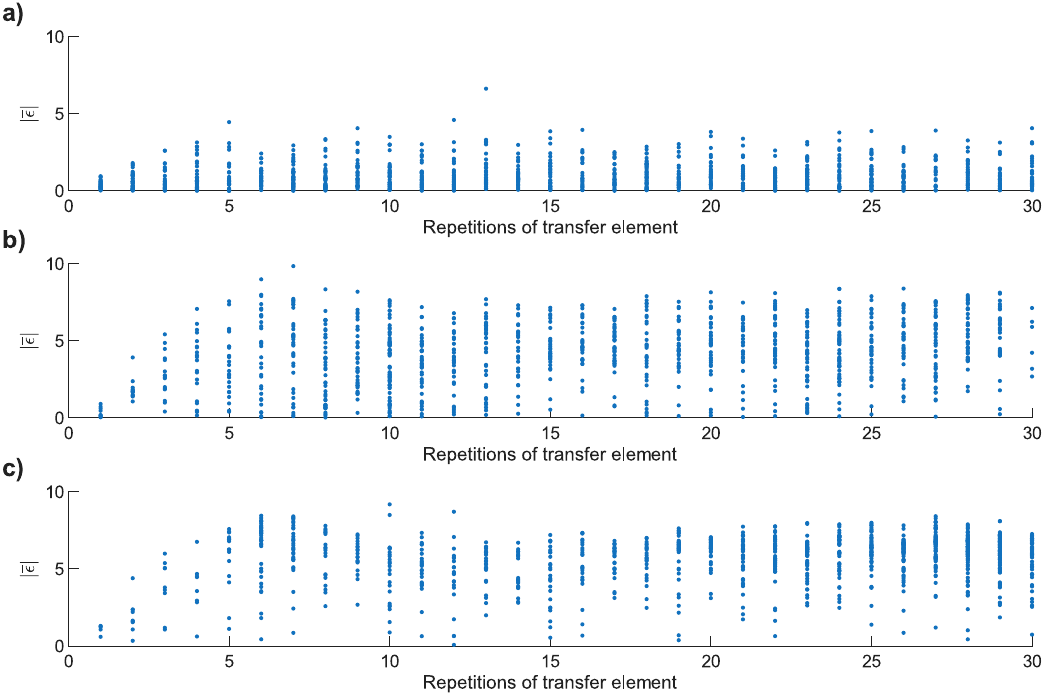} 
    \caption{Enhancement factors for the DNP transfer element pulse sequences found in Figs. 2 a1-c1  as a function of the number of times the DNP transfer element was repeated for the given sequence. The optimizations were, as specified in Fig. 2: a) Monte Carlo, b) Bayesian, and c) constrained Bayesian.}
    \label{fig:Fig2Rep}
\end{figure}

Pulse sequence optimizations involved:

\begin{itemize}
\item Fig. \ref{fig:OffsetProfiles-d5}: Experimental offset profiles recorded for  trityl OX063 in glycerol-d$_5$ for the Bayesian optimized DNP transfer pulse sequence in Fig. 2e relative to NOVEL, PLATO, and cRW-OPT1 DNP transfers (to compare with Fig. 2f using glycerol-d$_8$ in the solvent).

\item Fig. \ref{fig:Fig2EffField}: Analysis of the distribution of enhancement factors for the optimizations in Fig. 2 (main text) as function of the accumulated nutation angle $\theta^{DNP}$. The aim of this analysis is to explore the ability of \textit{in situ} Bayesian optimization to cope with the challenge of MW inhomogeneity which should favor good sequences with small accumulated nutation angles. Indeed, the best sequences in Fig. \ref{fig:Fig2EffField}a (Monte Carlo) and Fig. \ref{fig:Fig2EffField}b centers around resonances with $k_I$ values of 2 (ZQ and DQ) providing the highest transfer. Less efficient, but still highly populated, sequences are also found around the $k_I$=0 band. In particular the former illustrates (as also clear from the trajectories in Fig. 2, main text) that unconstrained Bayesian finds sequences coping with MW inhomogeneity. This is not effectively captured by the Monte Carlo optimizations. In the constrained optimizations, some flexibility around the resonance is observed.

\item Fig. \ref{fig:Fig2Rep}: Analysis of the distribution of enhancement factors for the optimizations in Fig. 2 (main text) as function of the number of repetitions of the DNP transfer element. The number of elements  was allowed to vary from 1 to 30 in the optimizations.

\item Fig. \ref{fig:PulseNumExp}:   Analysis of \textit{in situ} constrained cRW Bayesian optimizations on  trityl OX063 with glycerol-d$_5$ using different number of pulses in the 120 ns DNP transfer element.

\item Fig. \ref{fig:TEMPOopts}:  Experimental analysis of pulse sequence elements (excitation pulse and transfer element) for DNP on 2 mM 4-Oxo-TEMPO-d$_6$. The analysis explores \textit{in situ} Bayesian optimizations for  different pulse sequences, leading to the setup used in Fig. 4.

\item Fig. \ref{fig:TEMPOsweeps}:  Experimental optimization of contact time and MW amplitude for NOVEL-type constant amplitude spin-lock DNP for 2 mM 4-Oxo-TEMPO-d$_6$ relating to the experiments in Fig. 4. It is noted that highest transfer is obtained at higher MW amplitude than used for NOVEL (14.8 MHz) while good transfer is also obtained at the NOVEL condition.

\item Fig. \ref{fig:NMRvsDNP}: Experimental offset profile recorded for trityl OX063  in glycerol-d$_5$ for the Bayesian optimized 4-pulse excitation pulse followed by a cRW-OPT1 DNP transfer element (repeated 5) times  along with comparison of $^1$H NMR spectra recorded with and without DNP with the optimal 4-pulse-excitation-cRW-OPT1 DNP sequence.

\item Fig. \ref{fig:OffsetProfiles-d5_pulse}: Experimental offset profiles recorded for  trityl OX063 in glycerol-d$_5$ for the Bayesian optimized 4-pulse excitation pulse sequence combined with cRW-OPT1 (5 repetions) DNP transfer pulse sequence in Fig. 3b relative to NOVEL, PLATO, and cRW-OPT1 DNP transfers (to compare with Fig. 3f using glycerol-d$_8$ in the solvent).

\end{itemize}

\subsection{Optimized pulse sequences}
\label{S14}

The optimal pulse sequences obtained by \textit{in situ} Bayesian optimization is represented by DNP transfer sequences for trityl and TEMPO radicals and excitation pulse for trityl as listed below with reference to figures in the main text.

\subsubsection{Optimized DNP transfer pulse sequences}
\label{S14a}

The DNP transfer pulse sequences resulting from \textit{in situ} Bayesian optimization is given in \Cref{tab:tritylDNP,tab:TEMPODNP} as specified below

\begin{itemize}
\item Table \ref{tab:tritylDNP}: Pulse sequence resulting from constrained Bayesian optimization of a 24-pulse DNP transfer element ($t_m$ = 120 ns) obtained by optimization on  trityl OX063 in glycerol-d$_5$ as represented in Fig. 2e (main text). 

\item Table \ref{tab:TEMPODNP}: Pulse sequence resulting from constrained Bayesian optimization of a 8-pulse DNP transfer element ($t_m$ = 10 $\mu$s; each pulse 1250 ns) obtained by optimization on a 5 mM sample of 4-Oxo-TEMPO-d$_6$ as represented in Fig. 4b (main text). 

\end{itemize}

\subsubsection{Optimized excitation pulses}
\label{S14b}

\begin{itemize}
\item Table \ref{tab:tritylEX}: Optimal 4-pulse excitation pulse combined with cRW-OPT1 DNP transfer (repeated 5 times) optimized for  trityl OX063 in glycerol-d$_5$ from Fig. 3a. 
\end{itemize}

\subsection{Parameters for optimizations}
\label{S15}

The optimizations presented in this work, both \textit{in situ} (experimentally) and \textit{in silico} (numerically) involved setting of optimization parameters which are presented in a series of tables with description of their specific connection to the figures presented in main text and in Supplementary Figures.

\subsubsection{\textit{in situ} optimizations}
\label{S15a}

Parameters used for \textit{in situ} optimizations on  trityl OX063 in glycerol-d$_5$ and 2 mM 4-Oxo-TEMPO-d$_6$. In the list below we specify which sample (Trityl or TEMPO) and figure the parameters corresponds to with more details given in the specific table.

\begin{itemize}
\item Table \ref{tab:insito2ac}: Trityl, Fig. 2(a-c)
        
\item Table \ref{tab:Fig2e}: Trityl, Fig. 2e 

\item Table \ref{tab:insitoS13}: Trityl, Fig. \ref{fig:PulseNumExp}

\item Table \ref{tab:Fig3a}: Trityl, Fig.  3.a
     
\item Table \ref{tab:insito4S15}: TEMPO, Fig. \ref{fig:TEMPOopts}a,b

\item Table \ref{tab:lastoptions}: TEMPO, Fig. \ref{fig:TEMPOopts}c,d 

\item Table \ref{tab:insito4S15b}: TEMPO, Figs. 4 
\end{itemize}

\subsubsection{\textit{in silico} optimizations}
\label{S15b}

The numerical optimization used spin-pair parameters matching trityl with the tables presenting parameters used in optimizations relating to the marked figure.

\begin{figure}[ht]
    \centering
    \includegraphics[width=\linewidth]{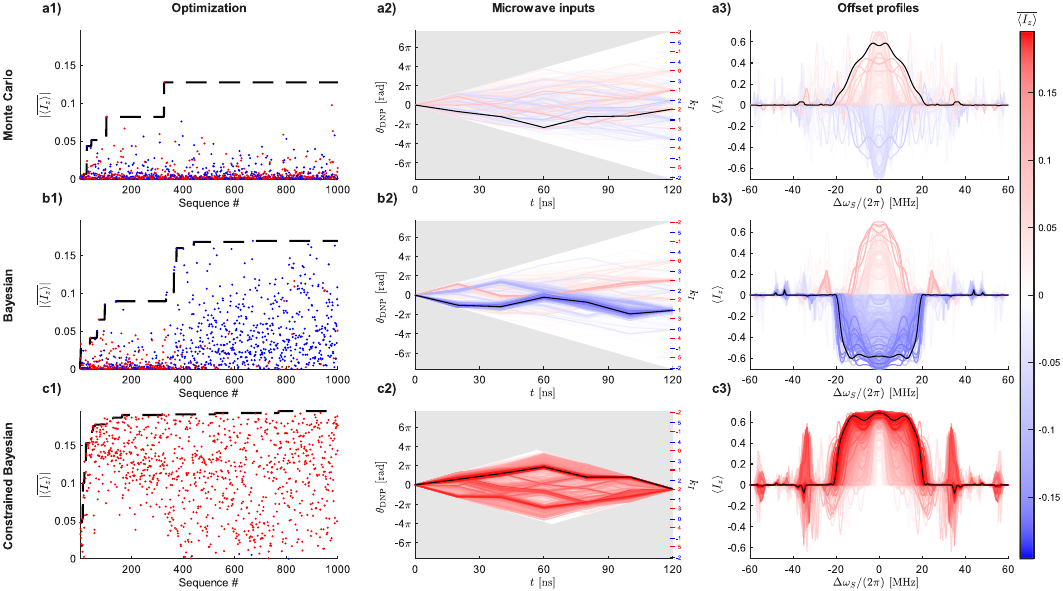} 
    \caption{Comparison of three different optimization methods by \textit{in silico} optimization: a) Monte Carlo, b) Bayesian, and c) constrained Bayesian. The layout is similar to Fig. 2(a-c). 
    (a1-c1) Absolute average polarization $|\overline{\langle I_z \rangle}|$  of the offset profile for each sequence, with red and blue dots representing match to ZQ and DQ resonances. (a2-c2) Accumulated nutation angle during transfer for each of the evaluated sequences. Colored numbers reflects $k_I$ values for ZQ (red) and DQ (blue). (a3-c3)  Offset profiles of the DNP polarization with the electron spin offset ranging from -60 to 60 MHz. In columns 2) and 3) the color is weighted by the absolute average enhancement. The best sequence in each optimization is marked in black. All results represent simulations for an electron-proton spin-pair with a distance of 4.5 Å, an initial density matrix $S_x$, a nuclear Larmor frequency 14.8 MHz, and a maximal MW nutation frequency of 32 MHz. A powder average was made using 20 polar angles between 0 and $\pi$ radians. Each pulse sequence element consisted of 6 $\pm x$-phase pulses each of duration of 20 ns, which were repeated until the most efficient transfer was reached. The acquisition function for the Bayesian optimization in b) and c) was "expected-improvement-plus". For the constraint in c) the cRW order "forward" was used with a DNP offset between $-5\degree$ and $5\degree$. Parameters for the optimizations are listed in Table \ref{tab:firstoptions}.}
    \label{fig:AcVar}
\end{figure}

\begin{figure}[ht]
    \centering
    \includegraphics[width=\linewidth]{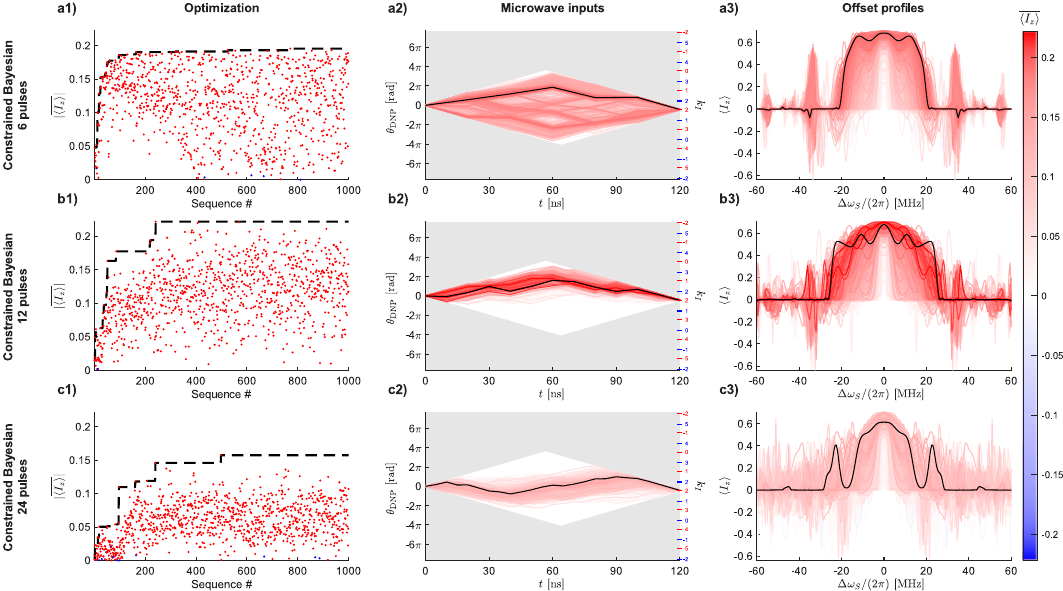} 
    \caption{The results from three \textit{in silico} constrained Bayesian optimizations using different numbers of pulses: a) 6 pulses, b) 12 pulses, and c) 24 pulses. In all cases, the modulation time is fixed to 120 ns. Otherwise, the layout and settings are the same as in Fig. \ref{fig:AcVar}c. Parameters for the optimizations are listed in Table \ref{tab:insilicoS7}.}
    \label{fig:PulseNum}
\end{figure}

\begin{figure}[ht]
    \centering
    \includegraphics[width=\linewidth]{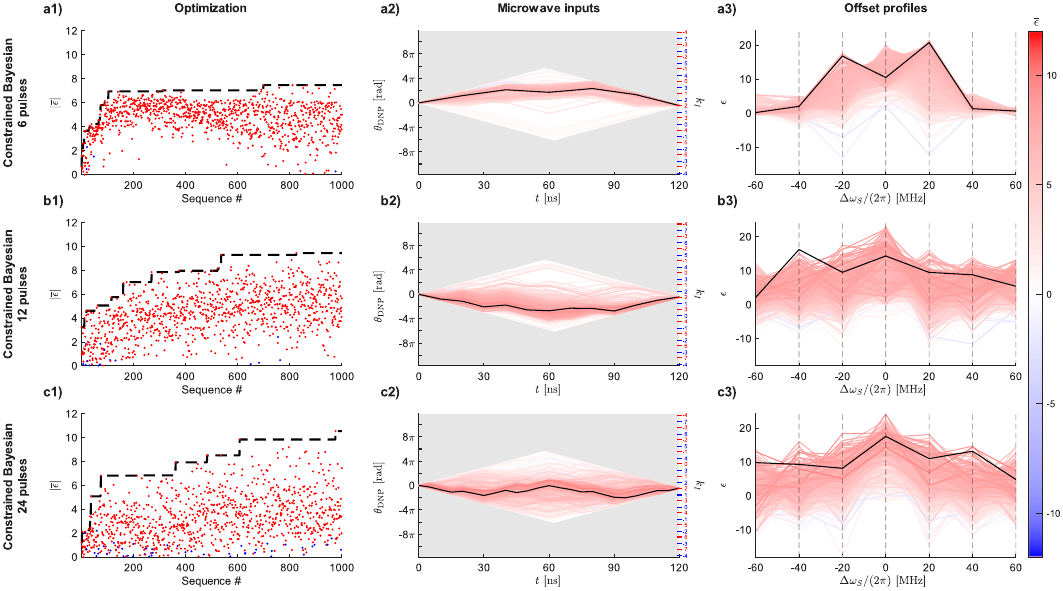} 
    \caption{Experimental \textit{in situ} constrained Bayesian optimizations of the DNP transfer for trityl OX063  with glycerol-d$_5$ using DNP transfer elements with different numbers of pulses: a) 6 pulses, b) 12 pulses, and c) 24 pulses. In all three cases, the modulation time is fixed at 120 ns. Here the constraint followed the "factorize" cRW order, but otherwise the layout and settings are the same as in Fig. 2c. Parameters for the optimizations are listed in Table \ref{tab:insitoS13}.}
    \label{fig:PulseNumExp}
\end{figure}

\begin{figure}[ht]
    \centering
    \includegraphics[width=\linewidth]{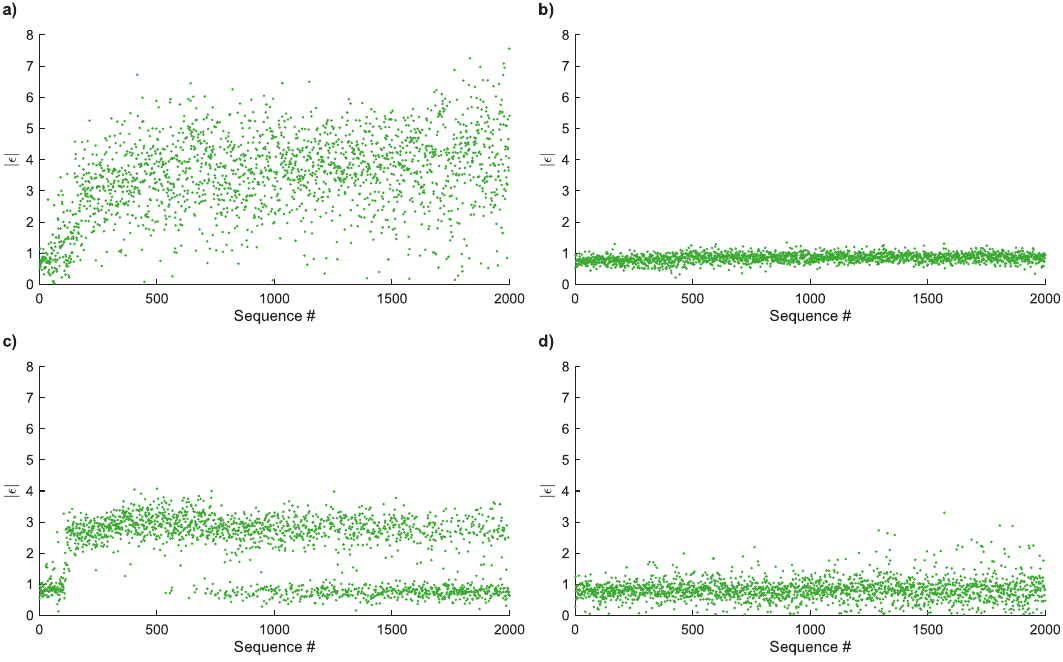} 
    \caption{\textit{in situ} (experimental) Bayesian optimization of DNP enhancement with 2 mM 4-Oxo TEMPO using different pulse sequence designs. a) 4-pulse excitation and 4-pulse DNP transfer element, both with arbitrary amplitude and phase. Excitation pulses had the same duration in the range of 0 - 5 ns. The DNP transfer element was allowed to be repeated 1 - 50 times, and the common duration of the transfer pulses was set to best match a total contact time of 10 $\mu$s. b) 20-pulse DNP transfer element with arbitrary amplitude and phase and no preceding excitation (i.e., aiming at longitudinal polarization transfer DNP \cite{carvalho2026}). The transfer element was repeated 1 - 100 times, and the common duration of the transfer pulses was set to best match a contact time of 10 $\mu$s. c) Fixed 52 MHz amplitude $y$-pulse excitation of duration of 5.5 ns combined with a 30-pulse DNP transfer element with arbitrary amplitude and $\pm x$ phase. The transfer element was repeated 1 - 100 times, and the common duration of the transfer pulses was set to best match a contact time of 10 $\mu$s. d) 4-pulse excitation and 10-pulse DNP transfer element with arbitrary amplitude and phase. Excitation pulses had the same duration in the range 0 - 5 ns. The transfer element was repeated 1 - 100 times, and the common duration of the transfer pulses was set to best match a contact time between 0 - 10 $\mu$s. Parameters for the optimizations are listed in Tables \ref{tab:insito4S15} and \ref{tab:lastoptions}.}
    \label{fig:TEMPOopts}
\end{figure}

\begin{figure}[ht]
    \centering
    \includegraphics[width=\linewidth]{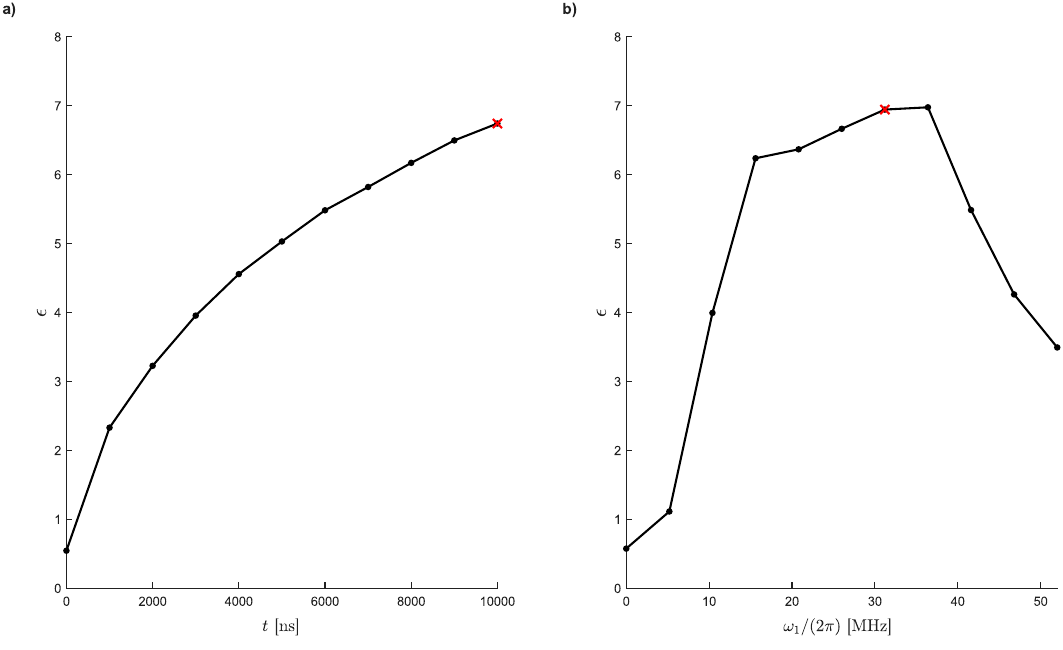} 
    \caption{\textit{in situ} (experimental) optimization of constant amplitude transfer using 4-Oxo-TEMPO with the red crosses marking the values used in Fig. 4: a) Contact time. b) MW nutation frequency.}
    \label{fig:TEMPOsweeps}
\end{figure}

\begin{figure}[ht]
    \centering
    \includegraphics[width=\linewidth]{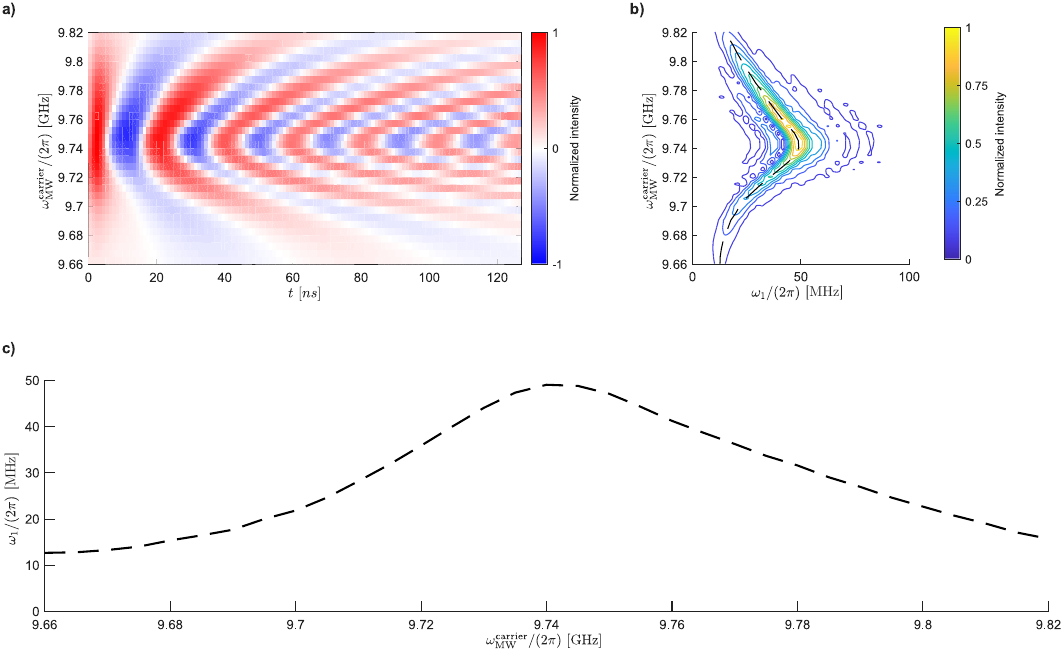} 
    \caption{a) Nutation curves for different MW carrier frequencies. The experiments used a 100\% amplitude pulse, in a digital scale, with the duration shown on the horizontal axis. The echo detection occured 10 $\mu$s after the start of the pulse, with pulse durations of 16 ns for the $\pi/2$-pulse and 32 ns for the $\pi$-pulse and a delay of 384 ns between them. The vertical axis indicates the carrier frequency of each nutation curve. To stay on resonance, the magnetic field followed the carrier $B_0=\omega^\text{carrier}_{MW}/\gamma_\text{exp}$ with a gyromagnetic ratio of $\gamma_\text{exp}=2.849$ MHz/G. b) Fourier transform of the nutation curves in a). The dashed line indicates the maximal intensity frequency for each carrier, which defines the resonator profile shown in c).}
    \label{fig:ResProf}
\end{figure}

\begin{figure}[ht]
    \centering
    \includegraphics[width=\linewidth]{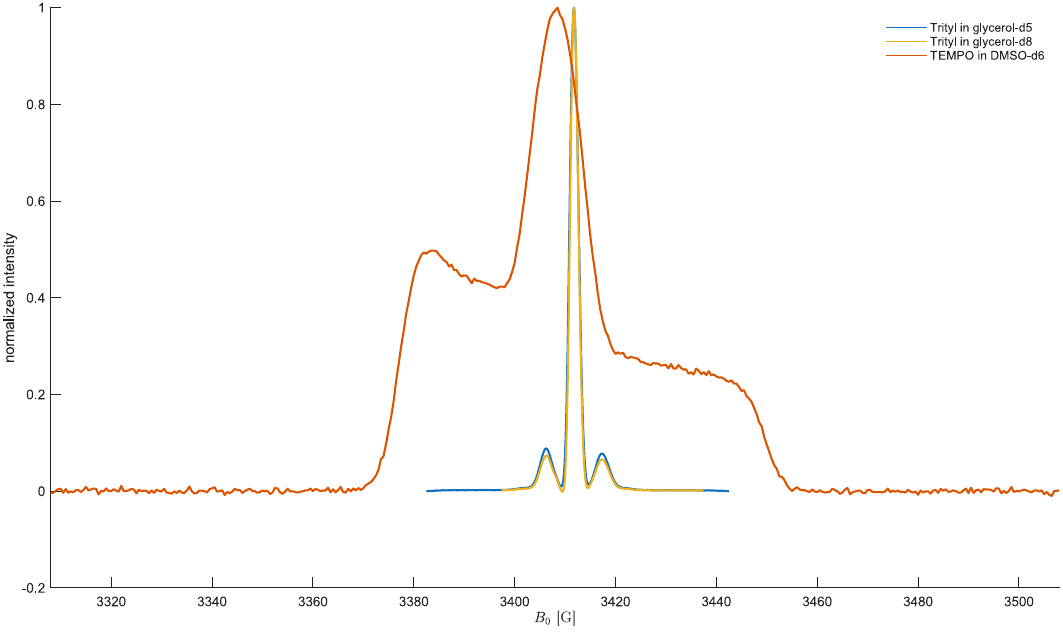} 
    \caption{Normalized EPR profiles for  trityl OX063 in glycerol-d$_5$ and glycerol-d$_8$ and 4-Oxo-TEMPO-d$_6$ obtained from field sweeps with the MW carrier at 9.72 GHz.}
    \label{fig:FieldSweeps}
\end{figure}

\begin{figure}[ht]
    \centering
    \includegraphics[width=\linewidth]{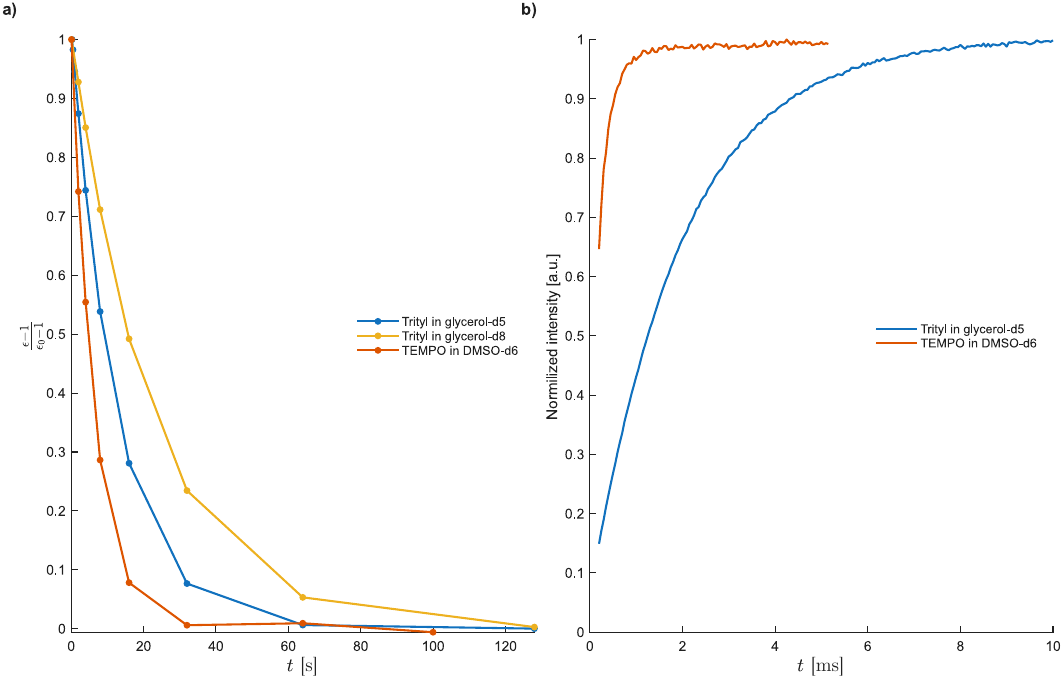} 
    \caption{a) Proton T$_1$ relaxation behavior after NOVEL DNP for the three different samples investigated in this study. The horizontal axis indicates the delay between DNP and the NMR solid echo and the vertical axis indicates the normalized enhancements after subtraction of the reference signal. Each sample uses a different normalization implying that the normalized enhancement is not comparable in-between different samples. For trityl in glycerol-d$_5$, the experiment used a 5.5 ns flip pulse of 45 MHz, and a 2 $\mu$s lock pulse of 19 MHz with a buildup of 64 s. For trityl in glycerol-d$_8$, the experiment used a 5.5 ns flip pulse of 40 MHz and a 8 $\mu$s lock pulse of 14.4 MHz with a buildup of 128 s. For TEMPO, the experiment used a 4 ns flip pulse of 52 MHz and a 10 $\mu$s lock pulse of 26 MHz with a buildup of 24 s. b) Measurement of electron recovery time as an indication of electron T$_1$ for samples of trityl and TEMPO. The vertical axis shows the normalized intensity of the EPR echo with the signal summed over 16 shots and the horizontal axis indicates the repetition time of each shot. For trityl, the experiment used 15 MHz echo pulses with a duration of 16 ns for the $\pi/2$-pulse and 32 ns for the $\pi$-pulse and a delay of 384 ns between them. For TEMPO, the experiment used 17 MHz echo pulses with duration 16 ns for the $\pi/2$-pulse and 32 ns for the $\pi$-pulse and a delay of 334 ns between them.}
    \label{fig:T1}
\end{figure}

\begin{figure}[ht]
    \centering
    \includegraphics[width=\linewidth]{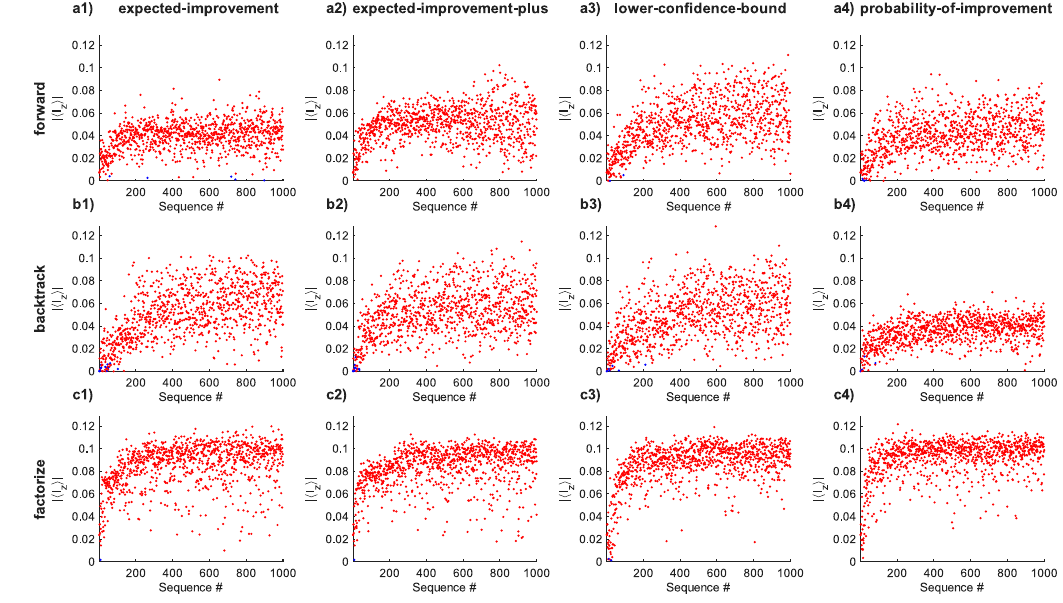}  
    \caption{\textit{in silico} analysis of different acquisition functions and cRW orders for a constrained Bayesian optimization of a 12-pulse DNP transfer element. The rows represent different ways of constructing the cRW: a) Accumulated nutation angles fixed in chronological order.  b) Accumulated nutation angles fixed in reverse order. c) Accumulated nutation angles most distant in time to the currently fixed angles, are fixed first. The columns represent different acquisition functions: 1) expected-improvement, 2) expected-improvement-plus, 3) lower-confidence-bound, and 4) probability-of-improvement.  Parameters for the optimizations are listed in   \Cref{tab:AmpAcFirst,tab:insilicoS5,tab:insilicoS75b,tab:AmpAcLast}.}
    \label{fig:AmpAc}
\end{figure}

\begin{figure}[ht]
    \centering \includegraphics[width=\linewidth]{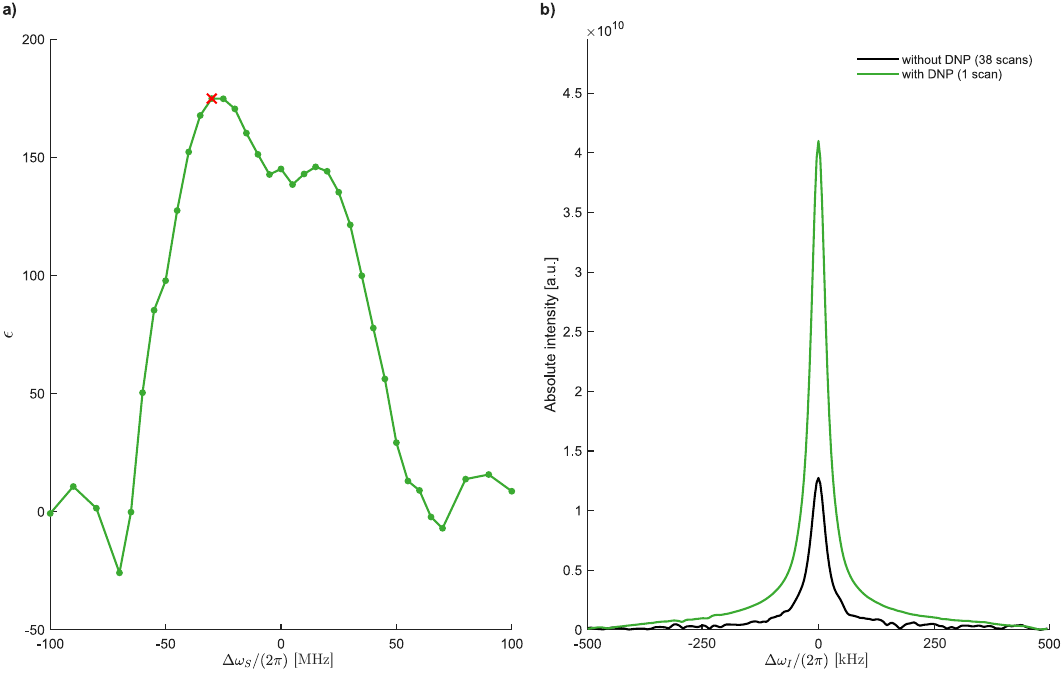}  
    \caption{a) Offset profile of the \textit{in situ} Bayesian optimized 4-pulse excitation pulse followed by a cRW-OPT1 DNP transfer element\cite{rCW_EEHT_DNP} repeated 5 times as in Fig. 3c. The profiles used the same pulse sequences employed in Fig. 3c but on trityl OX063 in glycerol-d$_5$ and with 5 s buildup time and were compared with corresponding profiles for NOVEL, PLATO, and cRW-OPT1. b) Comparison of $^1$H NMR spectra with (1 scan) and without (38 scans, 128 s repetition delay) DNP for  trityl OX063 in  glycerol-d$_8$ using the DNP experiment marked by a red cross in panel a). }
    \label{fig:NMRvsDNP}
\end{figure}

\begin{figure}[ht]
    \centering
    \includegraphics[width=0.55\linewidth]{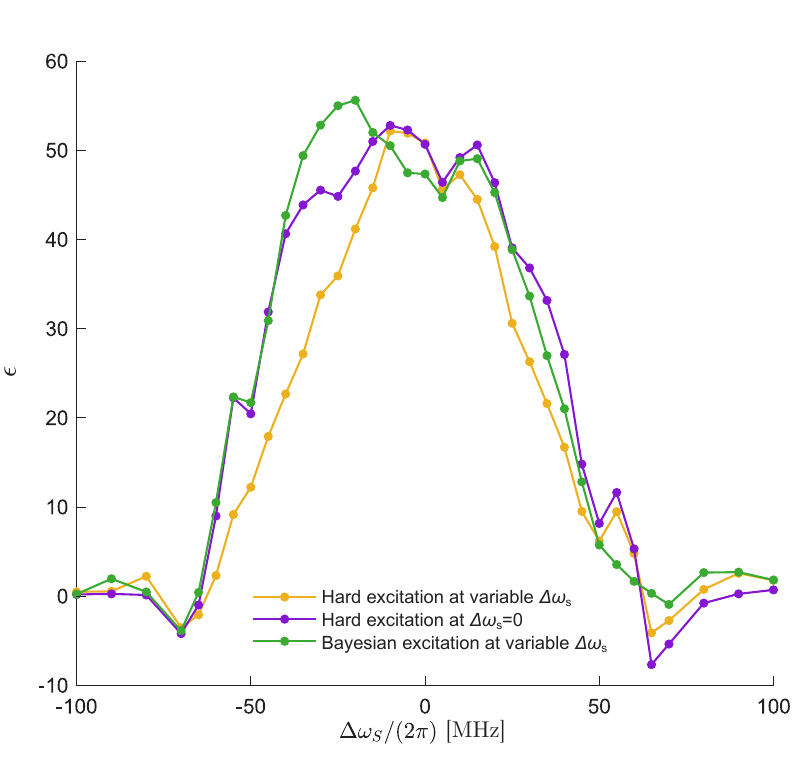} 
    \caption{Experimental \textit{in situ} polarization enhancements as a function of electron-spin offset using the same pulse sequences employed in Fig. 2f but on a sample of trityl in glycerol-d$_5$ and with 2 s buildup time.}
    \label{fig:OffsetProfiles-d5_pulse}
\end{figure}

\begin{itemize}

\item Table \ref{tab:firstoptions}: Fig. \ref{fig:AcVar}

\item Table \ref{tab:insilicoS7}:  Fig. \ref{fig:PulseNum}

\item Table \ref{tab:AmpAcFirst}: Fig. \ref{fig:AmpAc}(a1-c1)

\item Table \ref{tab:insilicoS5}: Fig. \ref{fig:AmpAc}(a2-c2)

\item Table \ref{tab:insilicoS75b}: Fig. \ref{fig:AmpAc}(a3-c3)

\item Table \ref{tab:AmpAcLast}: Fig. \ref{fig:AmpAc}(a4-c4)

\end{itemize}

\clearpage
\section{Supplementary Figures}
\label{S2}

\clearpage

\section{Supplementary Tables}
\label{S3}

\begin{table}[ht]
\footnotesize
\color{black}
\begin{tabular}{ l l l l}
\hline
Figure & 2 a1) & 2 b1) & 2 c1) \\ 
\hline
        Model & In-situ (Trityl) & In-situ (Trityl) & In-situ (Trityl) \\ 
        acFunc & monte-carlo & expected-improvement-plus & expected-improvement-plus \\ 
        deterministic & - & true & true \\
        B0 [T] & 3.4197 & 3.4197 & 3.4197 \\ 
        nuS [GHz] & 9.74 & 9.74 & 9.74 \\ 
        nu1max [MHz] & 49 & 49 & 49 \\ 
        nuI [MHz] & 14.795 & 14.795 & 14.795 \\ 
        NMRpulseTime [$\mu$s] & 1.1 & 1.1 & 1.1 \\ 
        eOffsets [MHz] & -60,-40,…,60 & -60,-40,…,60 & -60,-40,…,60 \\ 
        excitePulseNum & 1 & 1 & 1 \\ 
        exciteAmpScale & 0.96 & 0.96 & 0.96 \\ 
        excitePhase [rad] & $\pi/2$ & $\pi/2$ & $\pi/2$ \\ 
        excitePulseTime [ns] & 5 & 5 & 5 \\ 
        transPulseNum & 6 & 6 & 6 \\ 
        transAmpScale & -1 to 1 & -1 to 1 & -1 to 1 \\ 
        transPhase [rad] & 0 & 0 & 0 \\ 
        transPulseTime [ns] & 20 & 20 & 20 \\ 
        transRepeat & 1-30 & 1-30 & 1-30 \\ 
        RWorder & - & - & forward \\ 
        constrained & false & false & true \\ 
        DNPres & - & - & ZQ, $k_I=2$ \\ 
        DNPoffset [deg] & - & - & -5 to 5 \\ 
        buildup [s] & 2 & 2 & 2 \\ 
\hline
\end{tabular}
\caption{Overview of settings used in Figs. 2a-c. Each setting is explained in Supplementary Text Section \ref{sec:options}.}
\label{tab:insito2ac}
\end{table}

\begin{table}[ht]
\footnotesize
\color{black}
\begin{tabular}{ l l}
\hline
Figure  & 2 e) \\ 
        Model & In-situ (Trityl) \\ 
        \hline
        acFunc & expected-improvement-plus  \\ 
        deterministic & false  \\
        B0 [T] & 3.4142  \\ 
        nuS [GHz] & 9.72  \\ 
        nu1max [MHz] & 42  \\ 
        nuI [MHz]  & 14.7725 \\ 
        NMRpulseTime [$\mu$s] & 1.0  \\ 
        eOffsets [MHz] & -60,-40,…,60  \\ 
        excitePulseNum & 1 \\ 
        exciteAmpScale & 1  \\ 
        excitePhase [rad] & $\pi/2$  \\ 
        excitePulseTime [ns] & 5.25  \\ 
        transPulseNum & 24  \\ 
        transAmpScale & -1 to 1  \\ 
        transPhase [rad] & 0  \\ 
        transPulseTime [ns]  & 5  \\ 
        transRepeat & 6  \\ 
        RWorder & factorize \\ 
        constrained & true \\ 
        DNPres & ZQ, $k_I=2$ \\ 
        DNPoffset [deg] & -16.3 to 19.3  \\ 
        buildup [s] & 2 \\ 
\hline
\end{tabular}
\caption{Overview of settings used in Fig. 2e. Each setting is explained in Supplementary Text Section \ref{sec:options}.}
\label{tab:Fig2e}
\end{table}

\begin{table}[ht]
\footnotesize
\color{black}
\begin{tabular}{ l l l l}
    \hline
        Figure & \ref{fig:AcVar} a1) & \ref{fig:AcVar} b1) & \ref{fig:AcVar} c1) \\
        \hline
        Model & In-silico (2-spin) & In-silico (2-spin) & In-silico (2-spin) \\
        acFunc & monte-carlo & expected-improvement-plus & expected-improvement-plus \\
        deterministic & true & true & true \\
        nu1max [MHz] & 32 & 32 & 32 \\
        nuI  [MHz] & 14.8 & 14.8 & 14.8 \\
        eOffsets  [MHz] & -60,-59.5,…,60 & -60,-59.5,…,60 & -60,-59.5,…,60 \\
        excitation & ideal & ideal & ideal \\
        transPulseNum & 6 & 6 & 6 \\
        transAmpScale & -1 to 1 & -1 to 1 & -1 to 1 \\
        transPhase [rad] & 0 & 0 & 0 \\
        transPulseTime [ns] & 20 & 20 & 20 \\
        maxContact [ns] & 10e3 & 10e3 & 10e3 \\
        RWorder & - & - & forward \\
        constrained & false & false & true \\
        DNPres & - & - & ZQ, $k_I=2$ \\
        DNPoffset [deg] & - & - & -5 to 5 \\
        epDist [Å] & 4.5 & 4.5 & 4.5 \\
        polAngles & 20 & 20 & 20 \\
        MWinhom & false & false & false \\
        \hline
\end{tabular}
\caption{Overview of settings used in Fig. \ref{fig:AcVar}. Each setting is explained in Supplementary Text Section \ref{sec:options}.}
\label{tab:firstoptions}
\end{table}

\begin{table}[ht]
\footnotesize
\color{black}
\begin{tabular}{ l l l l}
\hline
        Figure & \ref{fig:PulseNum} a1) & \ref{fig:PulseNum} b1) & \ref{fig:PulseNum} c1) \\
        \hline
        Model & In-silico (2-spin) & In-silico (2-spin) & In-silico (2-spin) \\
        acFunc & expected-improvement-plus & expected-improvement-plus & expected-improvement-plus \\
        deterministic & true & true & true \\
        nu1max [MHz] & 32 & 32 & 32 \\
        nuI [MHz] & 14.8 & 14.8 & 14.8 \\
        eOffsets [MHz] & -60,-59.5,…,60 & -60,-59.5,…,60 & -60,-59.5,…,60 \\
        excitation & ideal & ideal & ideal \\
        transPulseNum & 6 & 12 & 24 \\
        transAmpScale & -1 to 1 & -1 to 1 & -1 to 1 \\
        transPhase [rad] & 0 & 0 & 0 \\
        transPulseTime [ns] & 20 & 10 & 5 \\
        maxContact [ns] & 10e3 & 10e3 & 10e3 \\
        RWorder & forward & forward & forward \\
        constrained & true & true & true \\
        DNPres & ZQ, $k_I=2$ & ZQ, $k_I=2$ & ZQ, $k_I=2$ \\
        DNPoffset [deg] & -5 to 5 & -5 to 5 & -5 to 5 \\
        epDist [Å] & 4.5 & 4.5 & 4.5 \\
        polAngles & 20 & 20 & 20 \\
        MWinhom & false & false & false \\
        \hline
\end{tabular}
\caption{Overview of settings used in Fig. \ref{fig:PulseNum}. Each setting is explained in Supplementary Text Section \ref{sec:options}.}
\label{tab:insilicoS7}
\end{table}

\begin{table}[ht]
\footnotesize
\color{black}
\begin{tabular}{ l l l l}
\hline
Figure & \ref{fig:PulseNumExp} a1) & \ref{fig:PulseNumExp} b1) & \ref{fig:PulseNumExp} c1) \\ 
                         \hline
        Model & In-situ (Trityl) & In-situ (Trityl) & In-situ (Trityl) \\ 
        acFunc & expected-improvement-plus & expected-improvement-plus & expected-improvement-plus \\ 
        deterministic & true & true & true \\
        B0 [T] & 3.4197 & 3.4197 & 3.4197 \\ 
        nuS [GHz] & 9.74 & 9.74 & 9.74 \\ 
        nu1max [MHz] & 49 & 49 & 49 \\ 
        nuI [MHz] & 14.795 & 14.795 & 14.795 \\ 
        NMRpulseTime [$\mu$s] & 1.1 & 1.1 & 1.1 \\ 
        eOffsets [MHz] & -60,-40,…,60 & -60,-40,…,60 & -60,-40,…,60 \\ 
        excitePulseNum & 1 & 1 & 1 \\ 
        exciteAmpScale & 0.96 & 0.96 & 0.96 \\ 
        excitePhase [rad] & $\pi/2$ & $\pi/2$ & $\pi/2$ \\ 
        excitePulseTime [ns] & 5 & 5 & 5 \\ 
        transPulseNum & 6 & 12 & 24 \\ 
        transAmpScale & -1 to 1 & -1 to 1 & -1 to 1 \\ 
        transPhase [rad] & 0 & 0 & 0 \\ 
        transPulseTime [ns] & 20 & 10 & 5 \\ 
        transRepeat & 1-30 & 1-30 & 1-30 \\ 
        RWorder & factorize & factorize & factorize \\ 
        constrained & true & true & true \\ 
        DNPres & ZQ, $k_I=2$ & ZQ, $k_I=2$ & ZQ, $k_I=2$ \\ 
        DNPoffset [deg] & -5 to 5 & -5 to 5 & -5 to 5 \\ 
        buildup [s] & 2 & 2 & 2 \\ 
\hline
\end{tabular}
\caption{Overview of settings used in Fig. \ref{fig:PulseNumExp}. Each setting is explained in Supplementary Text Section \ref{sec:options}.}
\label{tab:insitoS13}
\end{table}

\begin{table}[ht]
\footnotesize
\color{black}
\begin{tabular}{ c c c}
\hline
Pulse & $\omega_1/(2\pi)$ & $\Delta t$ \\
  & [MHz] & [ns] \\
\hline
1 & 10.5946 & 5 \\
2 & -30.2172 & 5 \\
3 & -17.0894 & 5 \\
4 & -8.83172 & 5 \\
5 & -41.2186 & 5 \\
6 & -37.1327 & 5 \\
7 & 39.1153 & 5 \\
8 & 25.697 & 5 \\
9 & -26.7359 & 5 \\
10 & 40.6605 & 5 \\
11 & 38.4145 & 5 \\
12 & 26.2706 & 5 \\
13 & -39.0101 & 5 \\
14 & -41.0916 & 5 \\
15 & -40.6515 & 5 \\
16 & -41.3883 & 5 \\
17 & -41.9525 & 5 \\
18 & -41.9024 & 5 \\
19 & 34.6693 & 5 \\
20 & 40.1102 & 5 \\
21 & 36.8637 & 5 \\
22 & 7.64619 & 5 \\
23 & 29.6807 & 5 \\
24 & 41.5037 & 5 \\
 \hline
\end{tabular}
\caption{DNP transfer sequence \#1918 from Fig. 2e. The pulses are $\pm x$-phase with the sign of the amplitude defining sign of the phase. The element was repeated 6 times.}
 \label{tab:tritylDNP}
\end{table}

\begin{table}[ht]
\footnotesize
\color{black}
\begin{tabular}{ c c c c}
\hline
Pulse & $\omega_1/(2\pi)$ & $\phi$ & $\Delta t$ \\
 & [MHz] & [Radians] & [ns] \\
\hline
1 & 5.79478 & 5.83871 & 3.375 \\
2 & 26.352 & 5.25783 & 3.375 \\
3 & 29.4872 & 1.9517 & 3.375 \\
4 & 31.8089 & 1.83484 & 3.375 \\
 \hline
\end{tabular}
\caption{DNP excitation sequence \#499 from Fig. 3a. cRW-OPT1 used $\pm x$-phase with the sign of the amplitude defining sign of the phase.}
 \label{tab:tritylEX}
\end{table}

\begin{table}[ht]
\footnotesize
\color{black}
\begin{tabular}{ l l}
\hline
Figure & 3 a)  \\ 
        Model & In-situ (Trityl)  \\ 
        \hline
        acFunc  & expected-improvement-plus \\ 
        deterministic  & false \\
        B0 [T]  & 3.4125 \\ 
        nuS [GHz]  & 9.72 \\ 
        nu1max [MHz]  & 45 \\ 
        nuI [MHz]   & 14.765 \\ 
        NMRpulseTime [$\mu$s]  & 0.95 \\ 
        eOffsets [MHz]  & -60,-40,…,60 \\ 
        excitePulseNum  & 4 \\ 
        exciteAmpScale  & 0 to 1 \\ 
        excitePhase [rad] & 0 to 2$\pi$ \\ 
        excitePulseTime [ns] & 1.25-5 \\ 
        transPulseNum  & 30 \\ 
        transAmpScale  & cRW-OPT1 \\ 
        transPhase [rad]  & 0 \\ 
        transPulseTime [ns]   & 5 \\ 
        transRepeat  & 5 \\ 
        RWorder  & forward \\ 
        constrained  & false \\ 
        DNPres & ZQ, $k_I=2$ \\ 
        DNPoffset [deg]  & 0.8 \\ 
        buildup [s]  & 5 \\ 
\hline
\end{tabular}
\caption{Overview of settings used in Fig. 3a. Each setting is explained in Supplementary Text Section \ref{sec:options}.}
\label{tab:Fig3a}
\end{table}

\begin{table}[ht]
\footnotesize
\color{black}
\begin{tabular}{ l l l}
\hline
Figure & \ref{fig:TEMPOopts} a) & \ref{fig:TEMPOopts} b)\\ 
        Model & In-situ (TEMPO) & In-situ (TEMPO) \\ 
        \hline
        acFunc & expected-improvement-plus & expected-improvement-plus\\ 
        deterministic  & false & false \\
        B0 [T]  & 3.408  & 3.408 \\ 
        nuS [GHz]  & 9.72 & 9.72\\ 
        nu1max [MHz] & 52& 52 \\ 
        nuI [MHz]  & 14.745 & 14.745 \\ 
        NMRpulseTime [$\mu$s]& 1.1 & 1.1  \\ 
        eOffsets [MHz]  & 0 & 0  \\ 
        excitePulseNum & 4  & 0  \\ 
        exciteAmpScale  & 0 to 1 & -\\ 
        excitePhase [rad] & 0 to $2\pi$ & -  \\ 
        excitePulseTime [ns] & 0 to 5 & - \\ 
        transPulseNum & 4 & 20\\ 
        transAmpScale & 0 to 1 & 0 to 1\\ 
        transPhase [rad]  & 0 to $2\pi$ & 0 to $2\pi$  \\ 
        transContact [ns] & 10e3  & 10e3 \\ 
        transRepeat & 1 to 50 & 1 to 100 \\ 
        constrained  & false & false  \\ 
        buildup [s] & 8  & 8 \\ 
\hline
\end{tabular}
\caption{Overview of settings used in Figs. \ref{fig:TEMPOopts}a-b. Each setting is explained in Supplementary Text Section \ref{sec:options}}
\label{tab:insito4S15}
\end{table}

\begin{table}[ht]
\footnotesize
\color{black}
\begin{tabular}{ l l l}
\hline
Figure & \ref{fig:TEMPOopts} c) & \ref{fig:TEMPOopts} d) \\ 
        Model & In-situ (TEMPO) & In-situ (TEMPO) \\ 
        \hline
        acFunc & expected-improvement-plus & expected-improvement-plus \\ 
        deterministic & false & false \\
        B0 [T] & 3.408 & 3.408 \\ 
        nuS [GHz]  & 9.72 & 9.72 \\ 
        nu1max [MHz] & 52 & 52 \\ 
        nuI [MHz] & 14.745 & 14.745 \\ 
        NMRpulseTime [$\mu$s] & 1.1 & 1.1 \\ 
        eOffsets [MHz] & 0 & 0 \\ 
        excitePulseNum & 1 & 4 \\ 
        exciteAmpScale & 1 & 0 to 1 \\ 
        excitePhase [rad] & $\pi/2$ & 0 to $2\pi$ \\ 
        excitePulseTime [ns] & 5.5 & 0 to 5 \\ 
        transPulseNum & 30 & 10 \\ 
        transAmpScale & -1 to 1 & 0 to 1 \\ 
        transPhase [rad] & 0 & 0 to $2\pi$ \\ 
        transContact [ns] & 10e3 & 0 to 10e3 \\ 
        transRepeat & 1 to 100 & 1 to 100 \\ 
        constrained & false & false \\ 
        buildup [s] & 8 & 8 \\ 
\hline
\end{tabular}
\caption{Overview of settings used in Figs. \ref{fig:TEMPOopts}c-d. Each setting is explained in Supplementary Text Section \ref{sec:options}.}
\label{tab:lastoptions}
\end{table}

\begin{table}[ht]
\footnotesize
\color{black}
\begin{tabular}{ c c c c}
\hline
Pulse & $\omega_1/(2\pi)$ & $\phi$ & $\Delta t$ \\  & [MHz] & [Radians] & [ns] \\
\hline
1 & 42.0768 & 0 & 4 \\
2 & 46.6575 & 4.43443 & 4 \\
3 & 47.5536 & 4.02352 & 4 \\
4 & 49.6457 & 3.08115 & 4 \\
5 & 25.9887 & 2.52414 & 1250 \\
6 & 23.6053 & 2.11097 & 1250 \\
7 & 9.33832 & 2.9136 & 1250 \\
8 & 27.4136 & 3.87421 & 1250 \\
9 & 21.8394 & 2.41519 & 1250 \\
10 & 23.3238 & 2.08272 & 1250 \\
11 & 42.2376 & 1.97648 & 1250 \\
12 & 22.9295 & 2.7669 & 1250 \\
 \hline
\end{tabular}
\caption{DNP sequence \#1021 from Fig. 4b. Transfer sequence was not repeated.}
 \label{tab:TEMPODNP}
\end{table}

\begin{table}[ht]
\footnotesize
\color{black}
\begin{tabular}{ l l l}
\hline
Figure & 4 a) & 4 b)  \\ 
        Model & In-situ (TEMPO) & In-situ (TEMPO)  \\ 
        \hline
        acFunc & monte-carlo & expected-improvement-plus  \\ 
        deterministic & - & false  \\
        B0 [T] & 3.408 & 3.408  \\ 
        nuS [GHz] & 9.72 & 9.72  \\ 
        nu1max [MHz] & 52 & 52  \\ 
        nuI [MHz] & 14.745 & 14.745  \\ 
        NMRpulseTime [$\mu$s] & 1.1 & 1.1  \\ 
        eOffsets [MHz] & 0 & 0  \\ 
        excitePulseNum & 4 & 4  \\ 
        exciteAmpScale & 0 to 1 & 0 to 1  \\ 
        excitePhase [rad] & 0 to $2\pi$ & 0 to $2\pi$  \\ 
        excitePulseTime [ns] & 0 to 5 & 0 to 5 \\ 
        transPulseNum & 8 & 8  \\ 
        transAmpScale & 0 to 1 & 0 to 1 \\ 
        transPhase [rad] & 0 to $2\pi$ & 0 to $2\pi$  \\ 
        transContact [ns] & 10e3 & 10e3  \\ 
        transRepeat & 1 & 1 \\ 
        constrained & false & false \\ 
        buildup [s] & 8 & 8 \\ 
\hline
\end{tabular}
\caption{Overview of settings used in Fig. 4. Each setting is explained in Supplementary Text Section \ref{sec:options}.}
\label{tab:insito4S15b}
\end{table}

\begin{table}[ht]
\footnotesize
\color{black}
\begin{tabular}{ l l l l}
\hline
        Figure & \ref{fig:AmpAc} a1) & \ref{fig:AmpAc} b1) & \ref{fig:AmpAc} c1) \\
                \hline
        Model & In-silico (2-spin) & In-silico (2-spin) & In-silico (2-spin) \\
        acFunc & expected-improvement & expected-improvement & expected-improvement \\
        deterministic & true & true & true \\
        nu1max [MHz] & 32 & 32 & 32 \\
        nuI [MHz] & 14.8 & 14.8 & 14.8 \\
        eOffsets [MHz] & -60,-59.5,…,60 & -60,-59.5,…,60 & -60,-59.5,…,60 \\
        excitation & ideal & ideal & ideal \\
        transPulseNum & 12 & 12 & 12 \\
        transAmpScale & -1 to 1 & -1 to 1 & -1 to 1 \\
        transPhase [rad] & 0 & 0 & 0 \\
        transPulseTime [ns] & 10 & 10 & 10 \\
        maxContact [ns] & 10e3 & 10e3 & 10e3 \\
        RWorder & forward & backtrack & factorize \\
        constrained & true & true & true \\
        DNPres & ZQ, $k_I=2$ & ZQ, $k_I=2$ & ZQ, $k_I=2$ \\
        DNPoffset [deg] & -5 to 5 & -5 to 5 & -5 to 5 \\
        epDist [Å] & 4.5 & 4.5 & 4.5 \\
        polAngles & 20 & 20 & 20 \\
        MWinhom & false & false & false \\
                \hline
\end{tabular}
\caption{Overview of settings used in the first column of Fig. \ref{fig:AmpAc}. Each setting is explained in Supplementary Text Section \ref{sec:options}.}
\label{tab:AmpAcFirst}
\end{table}

\begin{table}[ht]
\footnotesize
\color{black}
\begin{tabular}{ l l l l}
\hline
        Figure & \ref{fig:AmpAc} a2) & \ref{fig:AmpAc} b2) & \ref{fig:AmpAc} c2) \\
         \hline
        Model & In-silico (2-spin) & In-silico (2-spin) & In-silico (2-spin) \\
        acFunc & expected-improvement-plus & expected-improvement-plus & expected-improvement-plus \\
        deterministic & true & true & true \\
        nu1max [MHz] & 32 & 32 & 32 \\
        nuI [MHz] & 14.8 & 14.8 & 14.8 \\
        eOffsets [MHz] & -60,-59.5,…,60 & -60,-59.5,…,60 & -60,-59.5,…,60 \\
        excitation & ideal & ideal & ideal \\
        transPulseNum & 12 & 12 & 12 \\
        transAmpScale & -1 to 1 & -1 to 1 & -1 to 1 \\
        transPhase [rad] & 0 & 0 & 0 \\
        transPulseTime [ns] & 10 & 10 & 10 \\
        maxContact [ns] & 10e3 & 10e3 & 10e3 \\
        RWorder & forward & backtrack & factorize \\
        constrained & true & true & true \\
        DNPres & ZQ, $k_I=2$ & ZQ, $k_I=2$ & ZQ, $k_I=2$ \\
        DNPoffset [deg] & -5 to 5 & -5 to 5 & -5 to 5 \\
        epDist [Å] & 4.5 & 4.5 & 4.5 \\
        polAngles & 20 & 20 & 20 \\
        MWinhom & false & false & false \\
         \hline
\end{tabular}
\caption{Overview of settings used in the second column of Fig. \ref{fig:AmpAc}. Each setting is explained in Supplementary Text Section \ref{sec:options}.}
\label{tab:insilicoS5}
\end{table}

\begin{table}[ht]
\footnotesize
\color{black}
\begin{tabular}{ l l l l}
\hline
        Figure & \ref{fig:AmpAc} a3) & \ref{fig:AmpAc} b3) & \ref{fig:AmpAc} c3) \\
                 \hline
        Model & In-silico (2-spin) & In-silico (2-spin) & In-silico (2-spin) \\
        acFunc & lower-confidence-bound & lower-confidence-bound & lower-confidence-bound \\
        deterministic & true & true & true \\
        nu1max [MHz] & 32 & 32 & 32 \\
        nuI [MHz] & 14.8 & 14.8 & 14.8 \\
        eOffsets [MHz] & -60,-59.5,…,60 & -60,-59.5,…,60 & -60,-59.5,…,60 \\
        excitation & ideal & ideal & ideal \\
        transPulseNum & 12 & 12 & 12 \\
        transAmpScale & -1 to 1 & -1 to 1 & -1 to 1 \\
        transPhase [rad] & 0 & 0 & 0 \\
        transPulseTime [ns] & 10 & 10 & 10 \\
        maxContact [ns] & 10e3 & 10e3 & 10e3 \\
        RWorder & forward & backtrack & factorize \\
        constrained & true & true & true \\
        DNPres & ZQ, $k_I=2$ & ZQ, $k_I=2$ & ZQ, $k_I=2$ \\
        DNPoffset [deg] & -5 to 5 & -5 to 5 & -5 to 5 \\
        epDist [Å] & 4.5 & 4.5 & 4.5 \\
        polAngles & 20 & 20 & 20 \\
        MWinhom & false & false & false \\
                 \hline
\end{tabular}
\caption{Overview of settings used in the third column of Fig. \ref{fig:AmpAc}. Each setting is explained in Supplementary Text Section \ref{sec:options}.}
\label{tab:insilicoS75b}
\end{table}

\begin{table}[ht]
\footnotesize
\color{black}
\begin{tabular}{ l l l l}
\hline
        Figure & \ref{fig:AmpAc} a4) & \ref{fig:AmpAc} b4) & \ref{fig:AmpAc} c4) \\
                         \hline
        Model & In-silico (2-spin) & In-silico (2-spin) & In-silico (2-spin) \\
        acFunc & probability-of-improvement & probability-of-improvement & probability-of-improvement \\
        deterministic & true & true & true \\
        nu1max [MHz] & 32 & 32 & 32 \\
        nuI [MHz] & 14.8 & 14.8 & 14.8 \\
        eOffsets [MHz] & -60,-59.5,…,60 & -60,-59.5,…,60 & -60,-59.5,…,60 \\
        excitation & ideal & ideal & ideal \\
        transPulseNum & 12 & 12 & 12 \\
        transAmpScale & -1 to 1 & -1 to 1 & -1 to 1 \\
        transPhase [rad] & 0 & 0 & 0 \\
        transPulseTime [ns] & 10 & 10 & 10 \\
        maxContact [ns] & 10e3 & 10e3 & 10e3 \\
        RWorder & forward & backtrack & factorize \\
        constrained & true & true & true \\
        DNPres & ZQ, $k_I=2$ & ZQ, $k_I=2$ & ZQ, $k_I=2$ \\
        DNPoffset [deg] & -5 to 5 & -5 to 5 & -5 to 5 \\
        epDist [Å] & 4.5 & 4.5 & 4.5 \\
        polAngles & 20 & 20 & 20 \\
        MWinhom & false & false & false \\
                         \hline
\end{tabular}
\caption{Overview of settings used in the fourth column of Fig. \ref{fig:AmpAc}. Each setting is explained in Supplementary Text Section \ref{sec:options}.}
\label{tab:AmpAcLast}
\end{table}

\begin{table}[ht]
   \footnotesize
    \begin{tabular}{c c}
    \hline
         Amplitude factor & Weight \\
          & [\%] \\
         \hline
         0.65 & 4.84 \\
        0.70 & 5.29 \\
        0.75 & 5.86 \\
        0.80 & 6.62 \\
        0.85 & 7.69 \\
        0.90 & 9.39 \\
        0.95 & 12.73 \\
        1.00 & 32.88 \\
        1.05 & 14.68 \\
        \hline
    \end{tabular}
    \caption{The MW inhomogeneity profile used in Figs. 3e and 3f. For each row, the simulation using inhomogeneity averaging was run with the MW amplitudes multiplied by the amplitude factor.\cite{GUPTA201517} The weighted average was then taken using the specified weight from each row.}
    \label{tab:inhom}
\end{table}

\end{document}